\documentclass[prd,twocolumn,showkeys,preprintnumbers,nofootinbib,longbibliography
]{revtex4-1}

\usepackage{amsmath,amssymb,ifpdf,url,mathrsfs,slashed,multirow,tabularx,type1cm}
\usepackage{subfigure}
\usepackage{graphicx,color}
\usepackage{braket}
\usepackage{bm}
\usepackage[toc,page]{appendix}
\usepackage{comment} 

\allowdisplaybreaks

\definecolor{BlueViolet}{rgb}{0.2, 0.00, 0.7}
\definecolor{Blue}{rgb}{0.15, 0.00, 0.9}
\usepackage[
colorlinks=true,linkcolor=Blue,citecolor=Blue,
urlcolor=BlueViolet,breaklinks=true]{hyperref}

\newcommand{\Slash}[1]{{\ooalign{\hfil \hspace*{-5pt}~#1\hfil\crcr\raise.167ex\hbox{/}}}}

\def\({\left(}
\def\){\right)}
\def\<{\langle}
\def\>{\rangle}

\newcommand{\matl}{\left( \begin{array}}
\newcommand{\matr}{\end{array} \right)}

\def\beq#1\eeq{\begin{align}#1\end{align}}

\newcommand{\eg}{{\em e.g.}}

\begin{document}

\preprint{KEK--TH--2194}

\title{
Novel approach to neutron electric dipole moment search
using weak measurement
}

\author{Daiki Ueda}
\email{ueda@hep-th.phys.s.u-tokyo.ac.jp}
\affiliation{KEK Theory Center, IPNS, KEK, Tsukuba 305-0801, Japan,}
\affiliation{The Graduate University of Advanced Studies (Sokendai),
Tsukuba 305-0801, Japan}

\author{\vspace{-0.35cm}{\small and}\vspace{0.05cm}\\
Teppei Kitahara} \email{teppeik@kmi.nagoya-u.ac.jp} 
\affiliation{Physics Department, Technion---Israel Institute of Technology, Haifa 3200003, Israel,}
\affiliation{Institute for Advanced Research, Nagoya University, Nagoya 464-8601, Japan,}
\affiliation{Kobayashi-Maskawa Institute for the Origin of Particles and the Universe, Nagoya University, Nagoya 464-8602, Japan}

\date{\today}

\begin{abstract}
\noindent
We propose a novel approach in a search for the neutron electric dipole moment (EDM) by taking advantage of 
 signal amplification in a weak measurement, known as weak value amplification.
Considering an analogy to the weak measurement that can measure the spin magnetic moment interaction,
we examine an experimental setup with a polarized neutron beam through an external electric field with spatial gradient, 
where the signal is sensitive to the EDM interaction.
In particular,
a dedicated analysis of effects from impurities in pre- and post-selections is performed.
We show that the weak value amplification occurs 
where  the signal is enhanced by up to two orders of magnitude, and 
 demonstrate
 a potential sensitivity  of the proposed setup to the neutron EDM.
\end{abstract}


\maketitle 

\renewcommand{\thefootnote}{\#\arabic{footnote}}
\setcounter{footnote}{0}

\section{Introduction}
%
Since $CP$ violation arises from only the phase of the Cabibbo-Kobayashi-Maskawa matrix
in the standard model (SM) and it is tiny \cite{Kobayashi:1973fv}, 
$CP$-violating observables have provided good measurement sensitive to physics beyond the SM.
In particular, 
measurement of the electric dipole moment (EDM) of the neutron, $d_{\rm n}$,
can give a clear signal of new physics (NP), and 
has been a big subject for the last seventy years \cite{Ramsey:1982pq}. %

The neutron EDM arises from three-loop short-distance  \cite{Shabalin:1978rs,Khriplovich:1985jr,Czarnecki:1997bu},  two-loop long-distance  \cite{Khriplovich:1981ca,McKellar:1987tf}, 
one-loop contributions from the QCD theta term \cite{Dragos:2019oxn},
and  tree-level charm-quark contributions \cite{Mannel:2012qk} within the SM, 
while it can arise from one-loop diagrams in general NP models, 
 such as  multi-Higgs bosons \cite{Weinberg:1976hu,Deshpande:1976yp,Weinberg:1989dx,Barr:1990vd}, supersymmetric particles \cite{Ellis:1982tk,Buchmuller:1982ye,Polchinski:1983zd,Dugan:1984qf}, 
  leptoquark \cite{Arnold:2013cva,Fuyuto:2018scm, Dekens:2018bci},
  and models with dynamical electroweak symmetry breaking \cite{Appelquist:2004mn,Appelquist:2004es}.
In addition,
observed matter-antimatter asymmetry in the Universe requires 
new $CP$-violating sources \cite{Sakharov:1967dj,Gavela:1993ts}, which could be verified by the  measurement of the EDM, \eg, Ref.~\cite{Fuyuto:2015ida}.

So far, although much effort has been devoted to search for the EDMs, 
they have not been observed yet.
One of the most severe limits
comes from the neutron EDM search \cite{EDMex}
\begin{align}
(d_{\rm n})_{\rm exp}<3.0\times 10^{-26}~e\,{\rm cm}~(90\%~{\rm CL})\,, 
\label{eq:currentbound}
\end{align}
by measuring a neutron resonant frequency of  ultracold neutrons (UCNs)  based on
the separated oscillatory field method 
(the so-called Ramsey method) \cite{Ram1,Ram2}.\footnote{%
Very recently, 
an improved limit
has been announced by the nEDM collaboration \cite{newEDM,Abel:2020gbr}: 
\begin{align} (d_{\rm n})_{\rm exp}< 1.8 \times 10^{-26}~e\,{\rm cm}~(90\%~{\rm CL})\,.
\label{eq:currentbound_improved}
\end{align}}
This limit is five orders of magnitude larger than the SM prediction $(d_{\rm n})_{\rm SM}\sim 10^{-(31\textrm{--}32)}~e\,{\rm cm}$ 
\cite{Khriplovich:1981ca,McKellar:1987tf,Mannel:2012qk}.
Nevertheless, it severely constrains the NP scenarios that include additional $CP$ violation.

In the early stage of the neutron EDM experiments,
not the UCNs but a polarized neutron beam 
had been utilized  \cite{PhysRev:108:120,PhysRev:179:1285,PhysRevD:15:9,RAMSEY}. 
However, it was known that there was a large systematic uncertainty in neutron beam experiment which comes from relativistic effects.
The relativistic effects
arise from the motion of neutrons (velocity ${\bf v}$) through the electric field ${\bf E}$, as (see, \eg, Ref.~\cite{Feynman:2008ab} for a derivation)
\begin{align}
{\bf B} = \frac{{\bf E}\times {\bf v}}{c^2}\,.
\label{eq:relativistic}
\end{align}
Even if the neutron beam is shielded from the external magnetic field which we will assume in this paper, the external electric field \textit{does} generate the magnetic filed depending on the velocity (it can be  interpreted as the relativistic transformation of $F_{\mu\nu}$), and the sensitivity of the experiment becomes dull because of the large spin magnetic moment interaction.

In order to avoid large uncertainties,
current experiments 
and  new proposed projects 
are using the UCNs
\cite{Phys:at:nu59,PhysRevLett:97:131801,T:M:Ito, VANDERGRINTEN,A:P:Serebrov,S:K:Lam, C:Baker,PLA376,I:Alta}.
Moreover, most of the experiments 
have employed the Ramsey method 
\cite{Ram1,Ram2}.  
The main reasons why the UCNs are preferred are the following two
\cite{PhysRevC:88:045502}:   
First, 
the systematic uncertainty from the
relativistic effects
can be neglected because of its small velocity of the UCNs.
Second, 
the UCNs can have longer interaction times with the external electric field because they can be trapped easily, and the statistical uncertainty is suppressed.
%

\begin{figure*}[t]
\begin{center}
\includegraphics[width=18cm]{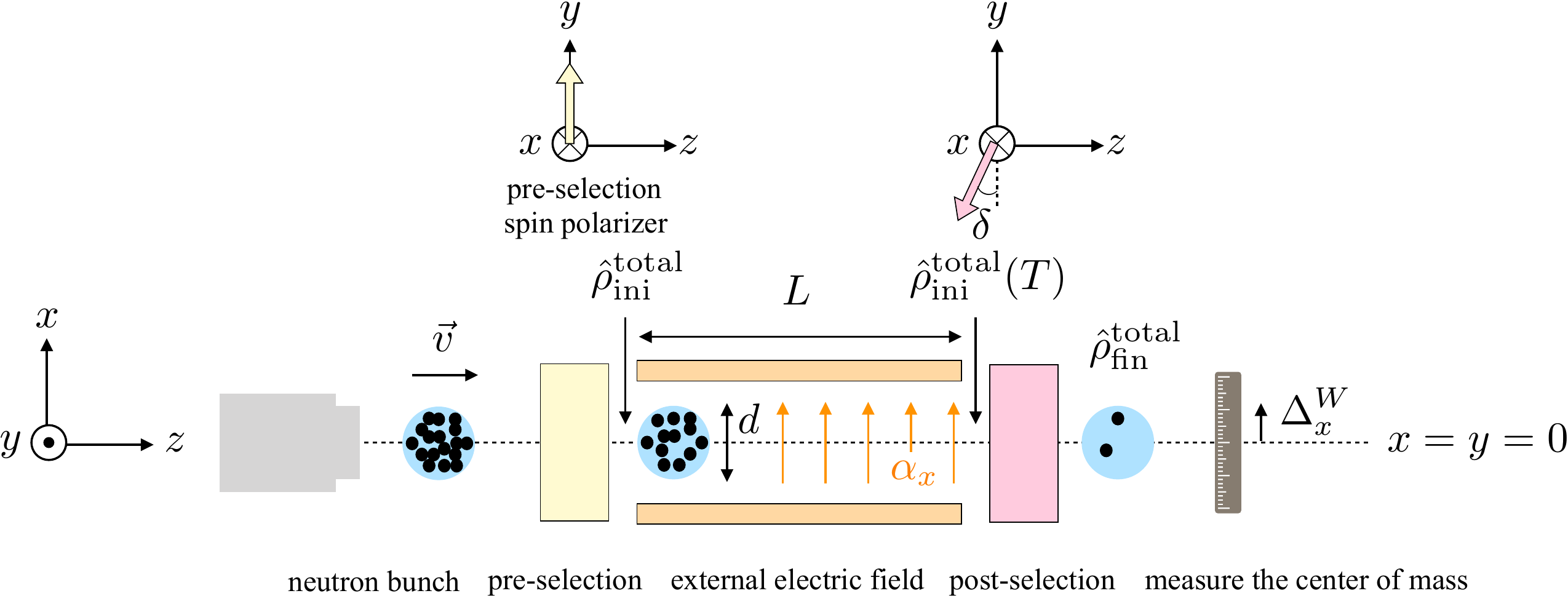}
\end{center}
%
\caption{
Schematic drawing of the proposed experimental setup.
First, emitted neutron bunches from neutron source (the leftmost gray box)
are polarized into the yellow arrow
in the first polarizer (the yellow box) and it is called pre-selection. 
The initial state in this paper is a neutron state 
that just goes through the first polarizer. 
After the pre-selection, 
the neutron bunches go through 
the external electric field with spatial gradient along $x$ axis for a distance $L \,(=v T)$. 
Then, 
the neutron bunches are selected by the second polarizer (the pink box) where a specific spin polarization state (the pink allow) can pass.
The final state is a neutron state
that just goes through the post-selection.
After the post-selection, the detector measures 
a position of the center-of-mass of the neutron bunch.
Spin directions of the two polarizers are exhibited above the polarizers.
The black dots in the neutron bunch
roughly
represent the single neutron:
the number of neutron significantly decreases after the post-selection.
}
\label{fig:setup}
\end{figure*}

On the other hand, 
in the polarized neutron beam experiments, 
although severe systematic effects come from the relativistic ${\bf E}\times {\bf v}$ corrections \cite{PhysRevD:15:9,PhysRevC:88:045502},
one can prepare a much larger amount of neutrons, 
which can reduce statistical fluctuations.
Also, one can use stronger external electric fields,
because the neutron beams are not covered by an insulating wall unlike the UCNs. 
Besides, there are several ideas that the relativistic ${\bf E}\times {\bf v}$ corrections can be suppressed in the neutron beam experiments with the Ramsey method  \cite{PhysRevC:88:045502} or with a spin rotation in non-centrosymmetric crystals \cite{FEDOROV200511,Fedorov:2008zza,FEDOROV2009124,FEDOROV2009538c}. 
The latter idea has been realized in an experiment \cite{Fedorov:2010sj}.

In this paper,
we propose a novel experimental approach in the search for  the neutron EDM by applying not the Ramsey method but a method of weak measurement  \cite{Aharonov_1964,Aharonov:1988xu,Lee_2014}, 
and discuss conditions 
how our setup  
can overtake
the current upper limit in Eq.~\eqref{eq:currentbound}.
{
This work is the first application of the weak measurement to the neutron EDM measurement.}
For instance, 
the spin magnetic moment interaction had been measured by the weak measurement  \cite{Aharonov:1988xu,Hasegawa}, 
where 
a tremendous amplification of signal (a component of a spin) emerged.
Also, the weak measurement using an
optical polarizer with a  laser beam has been realised  \cite{PhysRevLett.66.1107,PhysRevLett.94.220405}.
We also show that the relativistic ${\bf E}\times {\bf v}$ corrections are suppressed in this approach.

In the weak measurement, 
two quantum systems are prepared,
and then the initial and final states are properly selected
in one of the quantum systems, which are called pre- ($| \psi_i \rangle$) and post-selections ($ \langle \psi_f|$), respectively. 
A weak value,
corresponding to an observable $\hat{A}$,
is defined as 
\beq
\langle \hat{A}\rangle^W \equiv \frac{\langle \psi_f | \hat{A}| \psi_i \rangle }{ \langle \psi_f | \psi_i \rangle} \in \mathbb{C}\,,
\label{eq:weakvalue}
\eeq
and can be amplified by choosing the proper selections of the states: $\langle \psi_f | \psi_i \rangle \sim 0$, which is called weak value amplification (for reviews see, \eg, Refs.~\cite{doi:10.1063/1.3518209,2012PhR...520...43K}).
Since the weak value is obtained as an observable quantity corresponding to $\hat{A}$ in an intermediate measurement between 
$| \psi_i \rangle$ and $ \langle \psi_f|$
without disturbing the quantum systems,
measurement of the weak value plays an important role 
in the quantum mechanics itself. 
In addition, the weak value provides new methods for precise measurements 
 \cite{Aharonov_1991,Yokota_2009}.
In fact, 
the weak value amplification was applied to precise measurements such as the spin Hall effect of the light (four orders of magnitude amplified) \cite{Hosten787}
and the beam deflection in a Sagnac interferometer
(two orders of magnitude amplified) \cite{PhysRevLett.102.173601}. 

In our setup (see Fig.~\ref{fig:setup}), as  explained in details at the next section, 
 unlike the Ramsey method with using the UCN,
 we consider a polarized neutron beam with the velocity of $\sim10^3~{\rm m/sec}$.
We investigate the motion of a neutron bunch 
in the external electric field with spatial gradient, 
and apply methods of the weak measurement which leads to  amplification of the signal.

Very interestingly, it will be shown in our setup that
the systematic uncertainty from 
the relativistic ${\bf E}\times {\bf v}$ effect
can be irrelevant compared to a neutron EDM signal. 
This fact is expected as a new virtue of the weak measurement {because our finding implies that the weak measurement itself is useful in quantum systems 
to suppress 
the systematic uncertainty such as the relativistic effect.}

This paper is organized as follows.
 In Sec.~\ref{sec:2}, we propose an experimental setup for the neutron EDM search using the weak measurement.
 In Sec.~\ref{sec:3}, we analytically calculate an expected observable in this setup. Especially, the weak value is introduced. 
 Numerical results are  evaluated in Sec.~\ref{sec:Result}. We will show the weak value amplification and a potential sensitivity to the neutron EDM signal in this setup.
 Finally, Sec.~\ref{sec:conclusion} is devoted for the conclusions. 
 In Appendix~\ref{App:A}, a general setup of an external electric field is considered.
 In Appendix~\ref{sec:Full}, 
 a formalism of a full-order calculation of the  expected observable is provided.

\section{Experimental setup}
\label{sec:2}
%
We consider 
application of the weak measurement method  \cite{Aharonov:1988xu} to
the neutron EDM measurement.
Because Ref.~\cite{Aharonov:1988xu} utilizes an external magnetic field with a spatial gradient for measuring the spin magnetic moment,
one should use the polarized neutron beam 
and an external electric field with a spatial gradient.

In order to obtain a signal amplification in a weak measurement (weak value amplification),
two important selections are necessary:
the pre-selection and the post-selection.
The pre-selection is equivalent to preparation of  the initial state in the conventional quantum mechanics.
On the other hand, 
the post-selection is extraction of a specific quantum state at the late time \cite{Aharonov_1964}, 
and it makes to understand the weak value amplification difficult because of lack of counterparts in the conventional quantum mechanics.
In a nutshell, 
the role of the post-selection is filtering where
only events that the observable (such as the position of the neutron) takes a large value are collected.
Note that the weak value amplification originates from the quantum interference \cite{Duck:1989jx,Vaidman,PhysRevA.91.032116,Qin:2016}, so that classical filtering is not suitable.
In our setup, 
we impose selections of the spin polarization of the neutrons as the pre- and post-selections.

Figure~\ref{fig:setup} shows our proposed experimental setup.
The detailed explanation is given in the figure caption and the following paragraph, especially the electric field with the spatial gradient $\alpha_x$ is represented by the orange arrows.
We set $xyz$ axis as follows:
The neutrons fly along the $z$ axis.
The external electric field has the gradient along $x$ axis.  
A spin direction of the pre-selection at the first spin polarizer is $y$ axis. 

The entire process of the setup
is divided into four stages:
\begin{enumerate}
\item {\it Pre-selection.}
At $t=0$, 
the neutron bunch emitted from the neutron source is polarized at the first polarizer (the yellow box in Fig.~\ref{fig:setup}),
and the quantum state of the total system is represented by the following density matrices
\begin{align}
    \hat{\rho}^{\rm total}_{\rm ini}=\hat{\rho}^P_{\rm ini}\otimes \hat{\rho}^S_{\rm ini}\,,
\end{align}
where $\hat{\rho}^P_{\rm ini}$ is the  initial state of the neutron position and $\hat{\rho}^S_{\rm ini}$ is the pre-selected spin polarization.

\item {\it Time evolution in external electric field.}
For $0 < t < T$,
the neutron bunch goes through the external electric field with spatial gradient in Fig.~\ref{fig:setup},
where the total system evolves by the Hamiltonian Eq.~\eqref{eq: total hamiltonian} as
\begin{align}
    \hat{\rho}^{\rm total}_{\rm ini}(t)=e^{-i\hat{H}t}\hat{\rho}^{\rm total}_{\rm ini}e^{i\hat{H}t}\,.
\end{align}
Note that in this paper, although we do not use the natural units, 
$\hbar$ is discarded for simplicity.

\item {\it Post-selection.}
After the time evolution of the total system in the external electric field, 
at $t=T$, the neutrons are selected in the second polarizer (the pink box in Fig.~\ref{fig:setup}) where a specific spin polarization state ($\hat{\rho}^S_{\rm fin}$) of the neutron  can pass.
Then, the neutron position is represented by a density matrix
\begin{align}
    \hat{\rho}^P_{\rm fin}={\rm Tr}_S\left[\hat{\rho}^S_{\rm fin}\hat{\rho}^{\rm total}_{\rm ini}(T)\right]\,.
\end{align}

\item {\it Measurement of the center of mass.}
After the post-selection, position shifts of the neutron along $x$-axis are measured.%
\footnote{Even if the post-selection is provided by the Stern–Gerlach apparatus,
a neutron position shift occurs in $y$--$z$ plane, and a position shift along $x$-axis does not happen there.
}
Here, by taking the average of those shifts, one can measure a position shift of the center-of-mass of the neutrons passing the post-selection.
The expectation value of the position shift is expressed as
\begin{align}
    \Delta^W_x =\frac{{\rm Tr}_P\left[\hat{x}\hat{\rho}^P_{\rm fin}\right]}{{\rm Tr}_P\left[\hat{\rho}^P_{\rm fin}\right]}\,.
\end{align}

\end{enumerate}
The detail evaluations of these processes will be discussed in the next section.

The external electric field is set between the pre-selection and the post-selection. 
As discussed in Appendix~\ref{App:A},
one can define the $x$ axis as the direction of spatial gradient of the external electric field.
Therefore, $x$ dependence of $y$ and $z$ components of the external electric field 
is negligible without loss of generality.
According as the size of the neutron EDM and the spin polarization,
displacement of the neutron along $x$ axis occurs.
 We would like to maximize this displacement by the weak value amplification.

In this setup, the single neutron can be described by the non-relativistic Hamiltonian \cite{Posp}:
\begin{align}
\hat{H}=\frac{\hat{\bf p}^2}{2 m_{\rm n}}-d_{\rm n}\hat{\bf E}\cdot \hat{\boldsymbol{\sigma}}-\frac{\mu_{\rm n}}{m_{\rm n}c^2}(\hat{\bf E}\times \hat{\bf p})\cdot \hat{\boldsymbol{\sigma}} \,,\label{eq: total hamiltonian}
\end{align} 
where $\hat{\bf p}$ is the momentum operator of a neutron, $m_{\rm n}$ is the neutron mass,
$d_{\rm n}$ is the neutron EDM, $\mu_{\rm n}$ is the neutron magnetic moment,
$\hat{\bf E}$
is an operator of the external electric field vector, and $\hat{\boldsymbol{\sigma}}$ is the spin operator corresponding to the polarized neutron: the operation on spin-states is defined as $\hat{\sigma}_{i} \left|{\pm}_{i}\right>=\pm\left|{\pm}_i\right>~{\rm for~} i=x,y,z$.
This Hamiltonian is defined in the Hilbert space $\mathcal{H}=\mathcal{H}_P \otimes \mathcal{H}_S$, where 
$\mathcal{H}_P$ is the Hilbert space of the position ($P$) of the neutron, while
$\mathcal{H}_S$ is that of the spin ($S$) of the neutron.\footnote{One has to consider the free-falling neutrons in the earth.
However, we assume that $x$ axis is perpendicular to the direction of the gravity force, so that we can treat the free-falling effects of neutrons independently.
}
Note that although the external magnetic field is zero in the setup, the magnetic filed is generated by the relativistic effect in Eq.~\eqref{eq:relativistic}.
%

\section{Weak measurement}
\label{sec:3}
%
In this section, 
we derive an analytic formula of
an expectation value of deviation of the neutron position from $x=0$, which is shown as
$\Delta_x^W$ in Fig.~\ref{fig:setup}.

To analytically study our strategy 
for the neutron EDM search based on the weak value amplification,
we adopt two assumptions as follows:

\textit{Assumption\,1}: We consider the following external electric field operator:
\begin{align}
\hat{\bf E}=
\left(E_{x}^0+\alpha \hat{x},\ E_{y}^0,\ 0 \right)\,,
\label{eq:Eapp}
\end{align}
where $E_x^0,\, E_y^0$, and $\alpha$ are constants (namely $\alpha =  dE_x/dx$).%
\footnote{
By adding $-\alpha \hat{z}$ to $z$ component of $\hat{\bf E}$, the electric field satisfies the equation of motion $\nabla\cdot \hat{\bf E}=0$ in the vacuum.
Our formalism does not change by the $z$ component of $\hat{\bf E}$.}
$\hat{x}$ is an operator corresponding to the $x$ coordinate of the neutron. 
A necessary condition of this form is discussed in 
Appendix~\ref{App:A}.
Even if the $E_z$ component is nonzero, 
the effect is irrelevant in this setup.
Although $E_z$ generates $B_{x,y}$ via the relativistic effects in the third terms of Eq.~(\ref{eq: total hamiltonian}),
they are significantly suppressed by 
small neutron momenta, $v_x$ and $v_y$.

\textit{Assumption\,2}:
We consider the following neutron initial state  in the Hilbert space $\mathcal{H}_P$:
\beq
\hat{\rho}^P_{\rm ini}&\equiv\left|G_{p_{x0}}\otimes p_{y0}\otimes p_{z0}  \right>\left<G_{p_{x0}}\otimes p_{y0}\otimes p_{z0}\right|,\label{eq: iniz}
\eeq
where we defined as
\beq
\langle x | G_{p_{x0}} \rangle &=  \frac{1}{(2 \pi d^2)^{1/4}} e^{i p_{x0} \cdot x} e^{ - \frac{x^2}{4 d^2}} \,,\\
& \nonumber \\
\langle j | p_{j0}\rangle &=\frac{1}{\sqrt{2\pi}} e^{i p_{j0} \cdot j} \,, \quad \textrm{for~} j=y,\,z\,.
\eeq 
Here, $p_{x0},\, p_{y0}$ and $p_{z0}$ are the initial neutron momenta, and ${|G_{p_{x0}}\rangle},\, {|p_{y0}\rangle}$ and ${|p_{z0}\rangle}$ are the quantum states of $x,\,y$ and $z$ directions, respectively.
We are interested in the spatial displacement of the neutron along $x$ axis.
As explained below,
in the weak measurement, 
the expectation value of the neutron position depends on only variance of the distribution.
Therefore, 
we assume the Gaussian wave packet $|G\rangle$ as the a quantum state of $x$ direction, for simplicity
\cite{Ishikawa:2005zc,Ishikawa:2018koj}. 
In the distribution, 
we regard 
$d$ as a standard deviation of the neutron beam and assume that the neutron beam diameter is $2d$ for the $x$ direction.\footnote{%
We also considered more realistic distribution that the neutron beam is described as a mixed state which is a statistical ensemble of single neutron state.
Here, we assumed that both states can be described as the Gaussian distributions, and the standard deviation of the mixed state and single neutron state are represented by $d$ and $d_{\rm single}$, respectively.
We checked that  $d_{\rm single}$ contributions to the following analysis are numerically irrelevant, and all the results are sensitive to only $d$. This justifies our assumption\,2. 
}
On the other hand,
for $y$ and $z$ directions,  we assume the plane wave.

In addition to above assumptions,
we use several numerical approximations in this section. 
These approximation are reasonable when one takes input values which will be used in Sec.~\ref{sec:Result}.
Note that these numerical approximations are not used in the final plot of Sec.~\ref{sec:Result}.

Based on the assumption\,1 in Eq.~\eqref{eq:Eapp}, the Hamiltonian in Eq.~(\ref{eq: total hamiltonian}) can be expressed as
\begin{align}
    \hat{H} = & \frac{\hat{\bf p}^2}{2m_{\rm n}}-g_{\mu}\left[\chi\left(E_x^0 +\alpha \hat{x}\right) +E_y^0 \hat{n}_{p_z}\right]\otimes\hat{\sigma}_x\notag
    \\
    &-g_{\mu}\left[\chi E_y^0 - 
    \left(E_x^0 +\alpha \hat{x}\right)\hat{n}_{p_z} \right]\otimes\hat{\sigma}_y\notag
    \\
    &-g_{\mu}\left[\left(E_x^0 +\alpha \hat{x}\right)\hat{n}_{p_y}-E_{y}^0 \hat{n}_{p_x} \right]\otimes\hat{\sigma}_z
    ,\label{eq:simp}
\end{align}
where we defined
 $p_0\equiv \sqrt{p_{x0}^2+p_{y0}^2+p_{z0}^2}$,
$g_{\mu}\equiv \mu_{\rm n}p_0 / (m_{\rm n}c^2)$, $\chi\equiv d_{\rm n}/g_{\mu}$, and 
$\hat{n}_{p_i}\equiv \hat{p}_i/p_0$ ($i=x,\,y,\,z$)  for convenience in the following analysis.
All interactions are normalized by $g_{\mu}$, and the EDM interaction is represented as $\chi g_{\mu} $.
Note that $\chi$ is dimensionless real quantity.

\begin{widetext}
The time evolution operator is $e^{- i \hat{H} t}$. 
Using the Baker--Campbell--Hausdorff formula, and $[\hat{x},\hat{p}_x] = i$ and $[ \hat{x}, \hat{p}^2_x] =  2 i \hat{p}_x$, we obtain
\beq
\exp \left( - i  \hat{H} t \right) = 
\exp \left( - i \frac{\hat{\bf p}^2}{2m_{\rm n}} t \right) \exp \left[ - i  \left( \hat{H}_0 + \hat{H}_{\chi}\right) t+ \mathcal{O}\left( t^3  g^2_{\mu} \frac{\alpha^2}{m_{\rm n}}\,,
t^3 g^2_{\mu} \alpha  E_y^0 \frac{\hat{p}_x}{m_{\rm n}} \right) \right]\,,
\label{eq:H0Hchi}
\eeq
where the interaction Hamiltonian with the background ($\hat{H}_0$) and with the EDM  ($\hat{H}_\chi$) are
\begin{align}
&\hat{H}_0
    \equiv-g_{\mu}E_x^0\left[\frac{E_{y}^0}{E_x^0}\hat{n}_{p_z}\otimes \hat{\sigma}_x - \hat{n}_{p_z}\otimes \hat{\sigma}_y +\left(\hat{n}_{p_y}-\frac{E_y^0}{E_x^0} \hat{n}_{p_x} \right)\otimes \hat{\sigma}_z \right]+g_{\mu}\alpha \left(\hat{x}+\frac{t}{2 m_{\rm n}}\hat{p}_x\right)\left(\hat{n}_{p_z}\otimes \hat{\sigma}_y -\hat{n}_{p_y}\otimes \hat{\sigma}_z\right)\,,
      \label{eq:H0}
    \\
&\hat{H}_{\chi}\equiv
-\chi  g_{\mu} \left[ 
E_{x}^0  +  \alpha \left(\hat{x}+\frac{t}{2 m_{\rm n}}\hat{p}_x\right) \right]\otimes \hat{\sigma}_x -\chi  g_{\mu}  
E_{y}^0  \otimes \hat{\sigma}_y
\,.
\label{eq:Hchi1}
\end{align}
We have checked that the $\mathcal{O}(t^3)$ terms in Eq.~\eqref{eq:H0Hchi} are numerically negligible
in the following analysis.
Moreover,
the last term in Eq.~\eqref{eq:Hchi1} is totally screened by $g_{\mu} E_{x}^0 \hat{n}_{p_z} \otimes \hat{\sigma}_y$ in $\hat{H}_0$.

Then, we expand the interaction Hamiltonian by $\alpha \hat{x}$, and obtain the following analytic form:
\begin{align}
    \exp \left[ - i  \left( \hat{H}_0 + \hat{H}_{\chi}\right) t \right]= \hat{U}_0(t) +
    \alpha \left(\hat{x}+\frac{t}{2 m_{\rm n}}\hat{p}_x\right)\!
    \hat{U}_1(t) +\mathcal{O}\left(\alpha^2\left(\hat{x}+\frac{t}{2 m_{\rm n}}\hat{p}_x\right)^2\right)\,,
\end{align}
with
\begin{align}
    \hat{U}_0(t)=&
    I_2
\cos \left(g_{\mu}E_x^0 t\right)
+
  \begin{pmatrix}
  -i\frac{E_y^0}{E_x^0}n_{p_{x0}} + i n_{p_{y0}} & -n_{p_{z0}}  +i \frac{E_y^0}{E_x^0} n_{p_{z0}}  +i\chi \\
n_{p_{z0}}  +i \frac{E_y^0}{E_x^0} n_{p_{z0}}  +i \chi  &  +i \frac{E_y^0}{E_x^0}n_{p_{x0}} -i n_{p_{y0}}
\end{pmatrix}\sin \left(g_{\mu}E_{x}^0 t \right)
+\mathcal{O}\left(\left(\frac{E_y^0}{E_x^0}\right)^2 \right)\,,
\\
\hat{U}_1 (t)=&
    \begin{pmatrix}
-i\frac{E_y^0}{E_x^0}n_{p_{x0}}+in_{p_{y0}}    & i\frac{E_y^0}{E_x^0}n_{p_{z0}} -n_{p_{z0}}+i \chi \\
 i\frac{E_y^0}{E_x^0}n_{p_{z0}}+n_{p_{z0}}+i\chi &  i\frac{E_y^0}{E_x^0}n_{p_{x0}}-in_{p_{y0}} 
\end{pmatrix}
g_{\mu}t \cos \left( g_{\mu} E_x^0 t\right) \notag\\
&+
\begin{pmatrix}
i \frac{E_y^0}{ (E_x^0)^2}n_{p_{x0}} -g_{\mu}t  &
-i \frac{E_y^0}{ (E_x^0)^2}n_{p_{z0}} \\
-i \frac{E_y^0}{ (E_x^0)^2}n_{p_{z0}} & -i \frac{E_y^0}{ (E_x^0)^2}n_{p_{x0}} -g_{\mu}t
\end{pmatrix}
\sin \left( g_{\mu}E_x^0 t \right) +\mathcal{O}\left(\left(\frac{E_y^0}{E_x^0}\right)^2 \right)\,,
\end{align}
where $I_2$ is the $2\times 2$ unit matrix.
Hereafter, we assume $\hat{n}_{p_i}\to n_{p_{i0}} = p_{i0}/p_0~(i=x,\,y,\,z)$
and $n_{p_{x0}}^2 + n_{p_{y0}}^2+ n_{p_{z0}}^2 = 1$ for simplicity of calculations.
The higher-order terms 
$\mathcal{O}( (E_y^0/E_x^0)^2) $ are numerically irrelevant when $E_{y}^0 \ll E_{x}^0$.
The $\hat{U}_0(t)$ and $\hat{U}_1(t)$
satisfy the unitarity condition:
\beq
\hat{U}_0(t) \hat{U}_0(t)^{\dag} =
I_2\,,~~~
\hat{U}_1(t) \hat{U}_1(t)^{\dag} =
\left(g_{\mu} t\right)^2
I_2\,,~~~
\hat{U}_0(t) \hat{U}_1(t)^{\dag} + \hat{U}_1(t) \hat{U}_0(t)^{\dag} = 0\,,
\eeq
up to the following higher-order corrections:
\beq
\mathcal{O}\left(\chi
\frac{E_y^0}{E_x^0} ,\,
\chi^2,\,\left(\frac{E_y^0}{E_x^0}\right)^2 ,\,
n_{p_{x0}}^2,\,\frac{E_y^0}{E_x^0} n_{p_{x0}} n_{p_{y0}}  \right)\,.
\eeq
Now, we obtain the compact form of the time evolution operator,
\beq
\exp \left( - i \hat{H} t \right) \simeq  
\exp \left( - i  \frac{\hat{\bf p}^2}{2m_{\rm n}}t\right)
\left[ 
\hat{U}_0(t) +
    \alpha \left(\hat{x}+\frac{t}{2 m_{\rm n}}\hat{p}_x\right)\!
    \hat{U}_1(t)
    \right]\,.
\eeq
\end{widetext}

As mentioned previous section, the weak measurement  requires pre- and post-selections, and
we select the neutron spin polarization.
Since the neutron polarization rate is not perfect in practical spin polarizers,
we include an impurity effect  in the pre- and post-selections as mixed spin states of the neutron.
As will be shown later, the final result significantly depends on the impurity effect.
This is because neutron passing probability at the post-selection is sensitive to the impurity effect in this setup,
and large passing probability dulls the neutron position shift.
 We consider the following pre-selected state in the Hilbert space $\mathcal{H}_S$:
\begin{align}
\hat{\rho}^S_{\rm ini}&=(1-\epsilon){|\psi\rangle}{\langle\psi|}+\epsilon {|\phi\rangle}{\langle\phi|}\notag
\\
&= \frac{1}{2} \begin{pmatrix}
1 &
-i (1-2\epsilon) \\
i(1-2\epsilon) & 1
\end{pmatrix}\,,
\label{eq:rhoini}
\end{align}
where two polarization states are (see Fig.~\ref{fig:setup})
\begin{align}
    {|\psi\rangle}&={|+_y\rangle}=\frac{1}{\sqrt{2}}
    \begin{pmatrix}
i  \\
-1
\end{pmatrix}\,,
    \\
    {|\phi\rangle}&={|-_y\rangle}=\frac{1}{\sqrt{2}}
    \begin{pmatrix}
i  \\
1
\end{pmatrix}\,.
\end{align}
Here, ${|\pm_y\rangle}$ are eigenstates of the spin operator $\hat{\sigma}_y$, and $\epsilon$ $(0 < \epsilon\ll 1)$ stands for the selection impurity.
After the pre-selection, the quantum state of the total system at the initial time $t=0$ can be expressed as  direct-product $\hat{\rho}_{\rm ini}^{\rm total}=
\hat{\rho}^P_{\rm ini}\otimes \hat{\rho}^S_{\rm ini},$
where $\hat{\rho}^P_{\rm ini}$ is defined in Eq.~(\ref{eq: iniz}), which is the assumption\,2.

\begin{widetext}
The late-time quantum state of the total system at $t=T$ (see Fig.~\ref{fig:setup}) just before the post-selection is given as
\begin{align}
&\hat{\rho}_{\rm ini}^{\rm total}(T)=e^{-i\hat{H}T}\hat{\rho}_{\rm ini}^{\rm total} e^{i \hat{H}T}\nonumber \\
&=
e^{-i \frac{\hat{\bf p}^2}{2m_{\rm n}} T}
\left[ 
\hat{U}_0(T) +
    \alpha \left(\hat{x}+\frac{T}{2 m_{\rm n}}\hat{p}_x\right)\!
    \hat{U}_1(T)
    \right]
\hat{\rho}_{\rm ini}^{\rm total}
\left[ 
\hat{U}_0(T)^{\dag} +
    \alpha \left(\hat{x}+\frac{T}{2 m_{\rm n}}\hat{p}_x\right)\!
    \hat{U}_1(T)^{\dag}
    \right]
e^{i \frac{\hat{\bf p}^2}{2m_{\rm n}} T}\,.\label{eq: evsta}
\end{align}

Next, we consider the following post-selected state:
\beq
\hat{\rho}^S_{\rm fin}&=(1-\epsilon){|\phi_{\delta}\rangle}{\langle \phi_{\delta}|}+\epsilon {|\psi_{\delta}\rangle}{\langle \psi_{\delta}|}\notag\\
&=\frac{1}{2} \begin{pmatrix}
1+(1-2\epsilon)\sin\delta &
i (1-2\epsilon)\cos\delta \\
-i(1-2\epsilon)\cos\delta & 1-(1-2\epsilon)\sin\delta
\end{pmatrix}\,,
\label{eq:rhofin}
\eeq
with
\beq
{|\psi_{\delta}\rangle}&\equiv e^{i\frac{\delta}{2}\hat{\sigma}_x}{|+_y\rangle}=\frac{1}{\sqrt{2}}\begin{pmatrix}
    i\left(\cos \frac{\delta}{2}-\sin \frac{\delta}{2} \right) \\
    - \left(\cos \frac{\delta}{2}+\sin \frac{\delta}{2} \right)
    \end{pmatrix}\,,\notag\\
{|\phi_{\delta}\rangle}&\equiv e^{i\frac{\delta}{2}\hat{\sigma}_x}{|-_y\rangle}=\frac{1}{\sqrt{2}}\begin{pmatrix}
    i\left(\cos \frac{\delta}{2}+\sin \frac{\delta}{2} \right) \\
    \cos \frac{\delta}{2}-\sin \frac{\delta}{2} 
    \end{pmatrix}\,.
\eeq
Here, $\delta$ is a polarization angle around $x$-axis for the post-selection (see Fig.~\ref{fig:setup}).
It is known that small $\delta$ angle is  preferred for the weak value amplification \cite{Aharonov:1988xu}.
The selection impurity $\epsilon$
is also included in the post-selection, and we assume its quality is the same as the pre-selection, for simplicity. 
After the post-selection,
the final state of the total system is written as $\hat{\rho}_{\rm fin}^{\rm total}=\hat{\rho}^P_{\rm fin}\otimes \hat{\rho}^S_{\rm fin}.$
Using the late-time state in Eq.~\eqref{eq: evsta}, 
we obtain 
the neutron final state in the Hilbert space $\mathcal{H}_P$,
\begin{align}
\hat{\rho}^P_{\rm fin}\equiv&
{\rm Tr}_S \left[ \hat{\rho}_{\rm fin}^{\rm total}\right]
=
{\rm Tr}_S\left[\hat{\rho}^S_{\rm fin}\hat{\rho}^{\rm total}_{\rm ini}(T) \right]
\nonumber
\\
=&{\rm Tr}_S\left[\hat{\rho}^S_{\rm fin} \hat{U}_0 (T)\hat{\rho}^S_{\rm ini}\hat{U}_0(T)^{\dagger}\right]
e^{-i \frac{\hat{\bf p}^2}{2m_{\rm n}}T }  
\left\{
\hat{\rho}^P_{\rm ini} +
\chi g_{\mu}\alpha T\left[ W\hat{\rho}^P_{\rm ini}\left(\hat{x}+\frac{T}{2m_{\rm n}}\hat{p}_x\right)+ W^{\ast}\left(\hat{x}+\frac{T}{2m_{\rm n}}\hat{p}_x\right)\hat{\rho}^P_{\rm ini}\right]
\right\}e^{i\frac{\hat{\bf p}^2}{2m_{\rm n}} T }
\nonumber\\
&
+\mathcal{O}\left(\left(g_{\mu}\alpha T\right)^2\right)\,,
\label{eq:fullform}
\end{align}
where we defined the following dimensionless complex quantity $W$,
\begin{align}
   & W \equiv 
   \frac{1}{\chi g_{\mu}T}
   \frac{{\rm Tr}_S\left[\hat{\rho}^S_{\rm fin}\hat{U}_0(T)\hat{\rho}^S_{\rm ini}\hat{U}_1(T)^{\dagger}\right]}{{\rm Tr}_S\left[\hat{\rho}^S_{\rm fin}\hat{U}_0(T)\hat{\rho}^S_{\rm ini}\hat{U}_0(T)^{\dagger}\right]}\,.
   \label{eq: weakvalue}
\end{align}
The $W$ corresponds to the weak value.
For the third term of Eq.~\eqref{eq:fullform}, using Eqs.~\eqref{eq:rhoini} and \eqref{eq:rhofin}, we used
\beq
{\rm Tr}_S\left[\hat{\rho}^S_{\rm fin}\hat{U}_1(T)\hat{\rho}^S_{\rm ini}\hat{U}_0(T)^{\dagger}\right] &= 
{\rm Tr}_S
\left[\hat{U}_0(T)^{\ast}(\hat{\rho}^S_{\rm ini})^T
\hat{U}_1(T)^T
(\hat{\rho}^S_{\rm fin})^T \right]
\nonumber \\
&= 
{\rm Tr}_S
\left[\hat{\rho}^S_{\rm fin} \hat{U}_0(T)\hat{\rho}^S_{\rm ini}
\hat{U}_1(T)^\dag
\right]^{\ast}\,.
\eeq
Similarly, one can easily find
${\rm Tr}_S\left[\hat{\rho}^S_{\rm fin}\hat{U}_0(T)\hat{\rho}^S_{\rm ini}\hat{U}_0(T)^{\dagger}\right] = {\rm Tr}_S
\left[\hat{\rho}^S_{\rm fin} \hat{U}_0(T)\hat{\rho}^S_{\rm ini}
\hat{U}_0(T)^\dag
\right]^{\ast}=$ real.

In an ideal experimental setup limit, $E_y^0/E_x^0 \to 0$, $n_{p_{x0,y0}} \to 0$, and $\epsilon \to 0$, 
the weak value $W$ is expressed as
\begin{align}
    W& =-i\frac{\langle{\psi(T)|\hat{\sigma}_x|\phi_{\delta}\rangle}}{\langle{\psi(T)|\phi_{\delta}\rangle}}
    + \frac{i}{\chi} 
    \frac{\langle{\psi(T)|\hat{\sigma}_y|\phi_{\delta}\rangle}}{\langle{\psi(T)|\phi_{\delta}\rangle}}
    \\
    & = \frac{i}{\chi} 
    -  e^{- 2 i g_{\mu} E_x^0 T } \cot \left(\frac{\delta}{2}\right)+ \mathcal{O}(\chi)
    \,,\label{eq: wekv}
\end{align}
where we define
\beq 
{|\psi(T)\rangle}
&=  \hat{U}_0(T){|\psi\rangle} 
\nonumber 
\\
&= \left\{\cos \left(g_{\mu} E_x^0 T\right) I_2 + i \sin \left(g_{\mu} E_x^0 T\right) \left[ \chi \hat{\sigma}_x  -  \hat{\sigma}_y \right]\right\} {|\psi\rangle} 
\\
&= e^{- i g_{\mu} E_x^0 T 
} {|\psi\rangle}  
+ i \sin \left(g_{\mu} E_x^0 T\right) \chi \hat{\sigma}_x  {|\psi\rangle}  
\,.
\eeq
According to the definition of the weak value in Eq.~\eqref{eq:weakvalue}, 
we find $W= - i \langle \hat{\sigma}_x\rangle^W + \frac{i}{\chi} \langle \hat{\sigma}_y\rangle^W $ in the ideal experimental limit, and show that
$W$ is amplified by  $\cot (\delta /2)$ for small $\delta$ region   \cite{Aharonov:1988xu}.
One should note that 
since $W$ is always multiplied by $\chi $ in Eq.~\eqref{eq:fullform},
the first term in Eq.~\eqref{eq: wekv} is not singular 
in $\chi \to 0$ limit.
In other words, there is a contribution in Eq.~\eqref{eq:fullform} that is independent of $\chi$ (signal) and sensitive to the weak value $W$, especially $ \langle \hat{\sigma}_y\rangle^W$.
We will show that such a contribution corresponds to
a background effect (from the relativistic ${\bf E}\times {\bf v}$ effect).
It would be interesting possibility to measure the weak value from the background effect, 
even if one cannot measure the neutron EDM signal.
It is noteworthy that
proposed setup is valuable 
for not only the neutron EDM search 
but also the quantum mechanics itself.

In practical experimental setup, 
since $E_y^0\neq 0$, $n_{p_{x0,y0}}\neq 0$, and $\epsilon \neq 0$,
$\mathcal{O}(1/\chi)$ term survives in
$W$ that induces  $\chi$-independent contributions in Eq.~\eqref{eq:fullform}.
This means that the neutron magnetic moment, which should be $\chi$ independent, behaves as a background effect 
against the neutron EDM signal 
in the weak measurement.  

Using $\hat{\rho}^P_{\rm fin}$ in Eq.~\eqref{eq:fullform}, 
one can consider $\textrm{Tr}_P [ \hat{\rho}^P_{\rm fin}] $
and $\textrm{Tr}_P [\hat{x} \hat{\rho}^P_{\rm fin}] $ as follows:
\begin{align}
{\rm Tr}_P \left[ \hat{\rho}^P_{\rm fin}\right]=&
    {\rm Tr}\left[\hat{\rho}^{\rm total}_{\rm fin}\right] \nonumber \\
    =& {\rm Tr}_S \left[\hat{\rho}^S_{\rm fin}\hat{U}_0(T)\hat{\rho}^S_{\rm ini} \hat{U}_0^{\dagger}(T) \right]
    \left(
    1  + \chi g_{\mu} \alpha  \frac{p_{x0} }{m_{\rm n} } T^2 \textrm{Re}W 
    \right)
    +\mathcal{O}\left( (g_{\mu}\alpha T)^2\right)\,, \label{eq:trans}
    \\
    {\rm Tr}_P \left[ \hat{x} \hat{\rho}^P_{\rm fin}\right]=&
    {\rm Tr}\left[\hat{x}\hat{\rho}^{\rm total}_{\rm fin}\right]
    \nonumber \\
    =&  
    {\rm Tr}_S \left[\hat{\rho}^S_{\rm fin}\hat{U}_0(T)\hat{\rho}^S_{\rm ini} \hat{U}_0^{\dagger}(T) \right]
    \Biggl(
    {\rm Tr}_P \left[\hat{x} e^{-i \frac{\hat{p}_x^2}{2 m_{\rm n}}T} \hat{\rho}^P_{\rm ini}e^ {i\frac{\hat{p}_x^2}{2 m_{\rm n}} T} \right] 
    \nonumber \\
& +
    \chi g_{\mu} \alpha T \left\{
W {\rm Tr}_P 
\left[ \hat{x} 
e^{-i \frac{\hat{p}_x^2}{2 m_{\rm n}}T}
\hat{\rho}^P_{\rm ini} \left( \hat{x} + \frac{T}{2 m_{\rm n}} \hat{p}_x \right)
e^{i \frac{\hat{p}_x^2}{2 m_{\rm n}}T}
\right]  
+
W^{\ast} {\rm Tr}_P 
\left[ \hat{x} e^{-i \frac{\hat{p}_x^2}{2 m_{\rm n}}T} \left( \hat{x} + \frac{T}{2 m_{\rm n}} \hat{p}_x \right)\hat{\rho}^P_{\rm ini}
e^{i \frac{\hat{p}_x^2}{2 m_{\rm n}}T}
\right] \right\} \Biggr)
\nonumber \\
 & +\mathcal{O}\left((g_{\mu}\alpha T)^2 \right)
 \nonumber \\
    =&  
    {\rm Tr}_S \left[\hat{\rho}^S_{\rm fin}\hat{U}_0(T)\hat{\rho}^S_{\rm ini} \hat{U}_0^{\dagger}(T) \right]
    \left(
   \frac{p_{x0}}{m_{\rm n}} T
  +
   2 \chi g_{\mu} \alpha T  \left(d^2+\frac{1+ 4d^2 p_{x0}^2}{4d^2}\frac{T^2}{2 m_{\rm n}^2} \right) \textrm{Re}
W 
-\chi g_{\mu} \alpha   \frac{T^2}{2 m_{\rm n}}{\rm Im}W  \right)
 \nonumber \\
 & +\mathcal{O}\left((g_{\mu}\alpha T)^2 \right)\,.
 \label{eq:trans2}
\end{align}
Here, we used the following relations [the assumption\,2 in Eq.~\eqref{eq: iniz}]:
\begin{align}
{\rm Tr}_P \left[ \hat{\rho}^P_{\rm ini}
\right] &= 1\,,\quad 
{\rm Tr}_P \left[ \hat{x} \hat{\rho}^P_{\rm ini}
\right] = 0\,, \quad 
{\rm Tr}_P \left[ \hat{x}^2 \hat{\rho}^P_{\rm ini}
\right] = d^2\,, \nonumber \\
{\rm Tr}_P \left[ \hat{p}_x \hat{\rho}^P_{\rm ini}
\right] &= p_{x0}\,,\quad
{\rm Tr}_P \left[ \hat{p}_x^2 \hat{\rho}^P_{\rm ini}
\right] = \frac{1+4 d^2 p_{x0}^2}{4d^2}\,,\quad 
{\rm Tr}_P \left[ \hat{x} \hat{p}_x \hat{\rho}^P_{\rm ini}
\right] = \frac{i}{2}\,,
\end{align}
and
\beq
\left[ \hat{x}, \, e^{- i \frac{\hat{p}_x^2}{2 m_{\rm n}}T }\right]  =  \frac{T}{m_{\rm n}} \hat{p}_x e^{- i \frac{\hat{p}_x^2}{2 m_{\rm n}}T }\,,
\eeq
from $[ \hat{x}, \hat{p}^2_x] =  2 i \hat{p}_x$.
Note that the neutron passing probability at the post-selection is expressed by ${\rm Tr}\left[\hat{\rho}^{\rm total}_{\rm fin}\right]$. Using Eq.~\eqref{eq:trans}, we obtain
\beq
{\rm Tr}\left[\hat{\rho}^{\rm total}_{\rm fin}\right] \approx 2 \epsilon + \frac{1}{4} \delta^2 \ll 1\,.
\label{eq:passing}
\eeq
This corresponds to reduction of the neutron beam intensity after the post-selection.

Eventually, 
we obtain an expectation value 
of the position shift of the neutron after the post-selection as (see Fig.~\ref{fig:setup}),
\beq
\Delta_x^W &\equiv 
\frac{\textrm{Tr}_P [\hat{x} \hat{\rho}^P_{\rm fin} ]}{\textrm{Tr}_P [ \hat{\rho}^P_{\rm fin} ]}\nonumber \\
&=
   \frac{p_{x0}}{m_{\rm n}} T
 +
   2\chi g_{\mu} \alpha T  \left[    d^2
+\frac{1}{2d^2}\left(\frac{T}{2 m_{\rm n}}\right)^2 
       \right] \textrm{Re}
W 
 -\chi g_{\mu} \alpha   \frac{T^2}{2 m_{\rm n}}{\rm Im}W
 +\mathcal{O}\left((g_{\mu}\alpha T)^2 \right)\,.
 \label{eq:exp}
\eeq
Note that the $(T/2m_{\rm n})^2/2d^2$ term in the third term is numerically negligible. 
In the limit of $n_{p_{x0,y0}} \to 0$,
we obtain the following analytical formula of the
expected position shift:
\beq
\Delta_x^W =  \Delta_x^W({\rm EDM}) + \Delta_x^W({\rm BG})
+\mathcal{O}\left(\epsilon^2,\,\delta^2,\, \left(\frac{E_y^0}{E_x^0}\right)^2,\,(g_{\mu}\alpha T)^2 \right)\,,
\label{eq:DeltaxWapp}
\eeq
with
\beq
\Delta_x^W ({\rm EDM})=&
-  \chi g_{\mu} \alpha T d^2 \delta \frac{1 - 3 \epsilon  }
{2  \epsilon}\cos \left(2 g_{\mu} E_x^0 T \right)
+\chi \alpha d^2 \frac{1 - \epsilon }{  \epsilon } \frac{E_y^0}{\left(E_x^0\right)^2}
\sin\left( g_{\mu}E_x^0 T \right)
\left[ 2  g_{\mu}E_x^0 T \cos\left( g_{\mu}E_x^0 T\right) -\sin\left( g_{\mu}E_x^0 T\right)\right]
\,,
\label{eq:DeltaxWappEDM}
\\
\Delta_x^W ({\rm BG})=&
\alpha \frac{T}{ m_{\rm n}}\delta \frac{1-\epsilon}{8 \epsilon}\frac{E_y^0}{(E_x^0)^2}\sin^2 \left( g_{\mu} E_x^0 T\right)
- \alpha  d^2
\delta
 \frac{1 - 3 \epsilon  }
{4 \epsilon }  
\frac{E_y^0}{\left(E_x^0\right)^2} \left[ 2 g_{\mu} E_x^0  T \cos\left(2 g_{\mu} E_x^0 T\right)-\sin\left(2  g_{\mu}E_x^0 T\right)\right]   \,.
\label{eq:DeltaxWappBG} 
\eeq
Here, the $\Delta_x^W ({\rm EDM})$ corresponds to the EDM signal, while the $\Delta_x^W ({\rm BG})$ is the shift by the background effect which stems from the neutron magnetic moment (the relativistic ${\bf E}\times {\bf v}$ effect).
The relativistic effects $\Delta_x^W ({\rm BG})$
mimic the neutron EDM signal $\Delta_x^W ({\rm EDM})$.

Surprisingly, 
we find that such the relativistic effect
is dropped in small $T$ region when $E_y^0 \ll E_x^0$ and/or $\delta \approx 0$.
For instance, when one takes $\delta =0$
in the post-selection, 
the expected position shift is
\begin{align}
    \Delta_x^W (\rm EDM)|_{\delta \to 0}&=\chi \alpha d^2 \frac{1 - \epsilon }{  \epsilon } \frac{E_y^0}{\left(E_x^0\right)^2}
\sin\left( g_{\mu}E_x^0 T \right)
\left[ 2 E_x^0 g_{\mu}T \cos\left( g_{\mu}E_x^0 T\right) -\sin\left( g_{\mu}E_x^0 T\right)\right]\,, \nonumber \\
    \Delta_x^W (\rm BG)|_{\delta \to 0}&=0\,.
\label{eq: ledel0}
\end{align}
In this limit which one can realize by setting the first and second polarizers to be turned the opposite directions,
the background shift from the relativistic ${\bf E}\times {\bf v}$ effect is dropped.
Note that for large $T$ region, the expansion of $\hat{\rho}^P_{\rm fin}$ with respect to $g_{\mu}\alpha T$ does not work, and suppression of $\Delta_x^W (\rm BG)$ no longer occurs.

 We observe that a dimensionless combination $g_{\mu} E_x^0 T $  is given by
\beq
g_{\mu} E_x^0 T \simeq -1.0\times 10^{-1} \left(\frac{v_{\rm n}}{10^3\, {\rm m}\cdot {\rm sec}^{-1}}\right) \left( \frac{E_x^0}{10^7\,{\rm V}\cdot {\rm m}^{-1}} \right) \left(\frac{T}{ 10^{-2} \,{\rm sec}}\right)\,,
\label{eq:gET}
\eeq
where we define the neutron velocity $v_{\rm n}$ as $p_0 = m_{\rm n} v_{\rm n}$. 
For $|g_{\mu} E_x^0 T |\ll 1$
region, 
the expected position shift in Eqs.~\eqref{eq:DeltaxWappEDM} and \eqref{eq:DeltaxWappBG} is given by
\beq
\Delta_x^W ({\rm EDM})\simeq & -  \chi g_{\mu} \alpha T d^2 \delta \frac{1 - 3 \epsilon  }
{2  \epsilon}
\left[1 - 2(  g_{\mu} E_x^0  T  )^2\right]
+\chi \alpha d^2 \frac{1 - \epsilon }{  \epsilon } \frac{E_y^0}{\left(E_x^0\right)^2}
\left( g_{\mu}E_x^0 T \right)^2
\,,
\\
\Delta_x^W ({\rm BG})\simeq &
\alpha \frac{T}{ m_{\rm n}}\delta \frac{1-\epsilon}{8 \epsilon}\frac{E_y^0}{(E_x^0)^2}\left( g_{\mu} E_x^0 T\right)^2
+2  \alpha  d^2
\delta
 \frac{1 - 3 \epsilon  }
{3 \epsilon }  
\frac{E_y^0}{\left(E_x^0\right)^2} (  g_{\mu} E_x^0  T  )^3  \,.
\label{eq:DeltaBG}
\eeq
Using $|E_{y}^0/E_{x}^0| \ll 1$ and $\epsilon \ll 1$, 
eventually we obtain an approximation formula,
\beq
\Delta_x^W \approx \Delta_x^W ({\rm EDM})\approx
-    \frac{\chi g_{\mu} \alpha T d^2 \delta} 
{2  \epsilon} = - d_{\rm n}  \frac{ \alpha T d^2 \delta} 
{2  \epsilon}
\,.
\eeq
\end{widetext}

\begin{figure*}[t]
\begin{center}
\includegraphics[width=8cm]{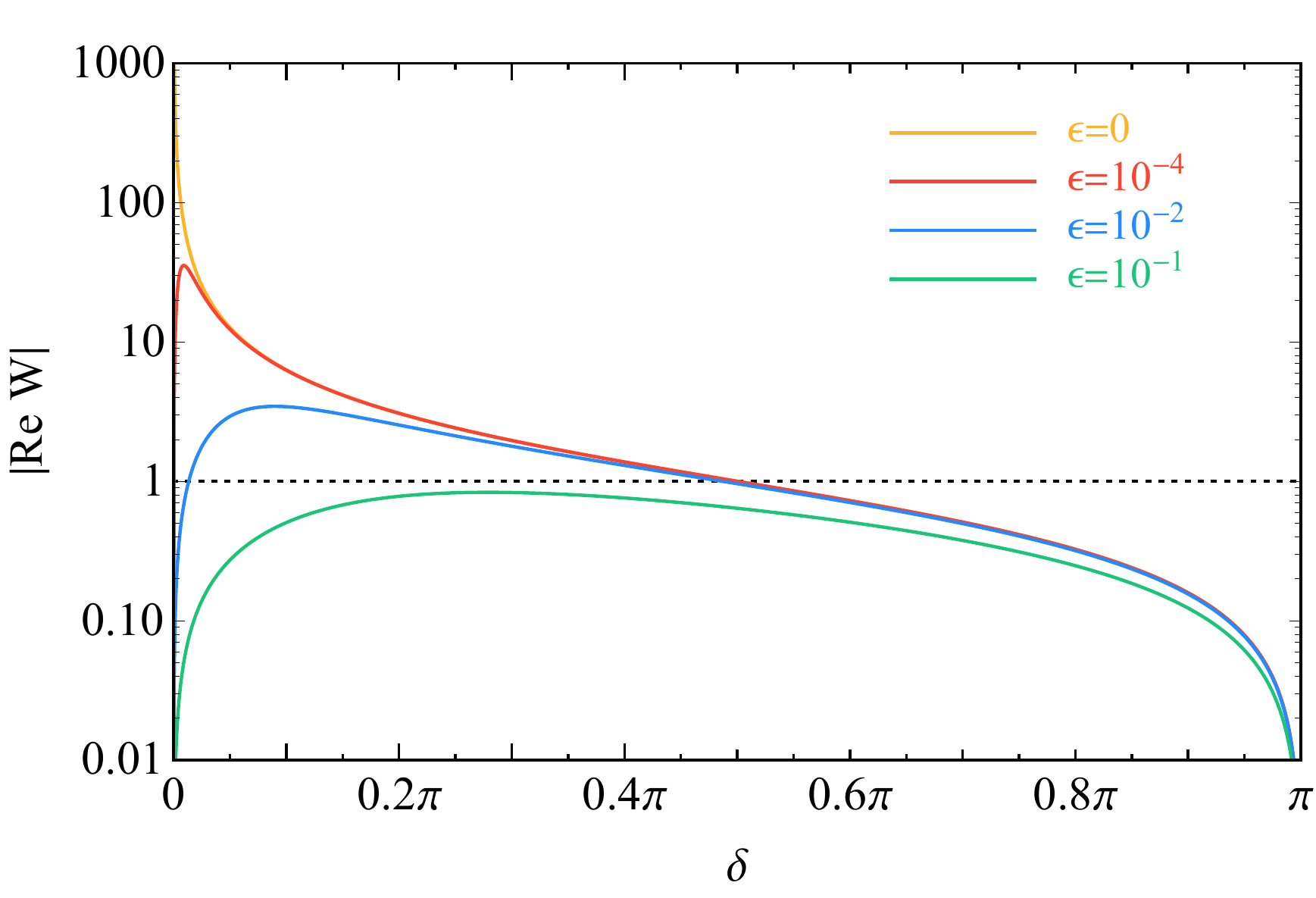}
\quad
\includegraphics[width=8cm]{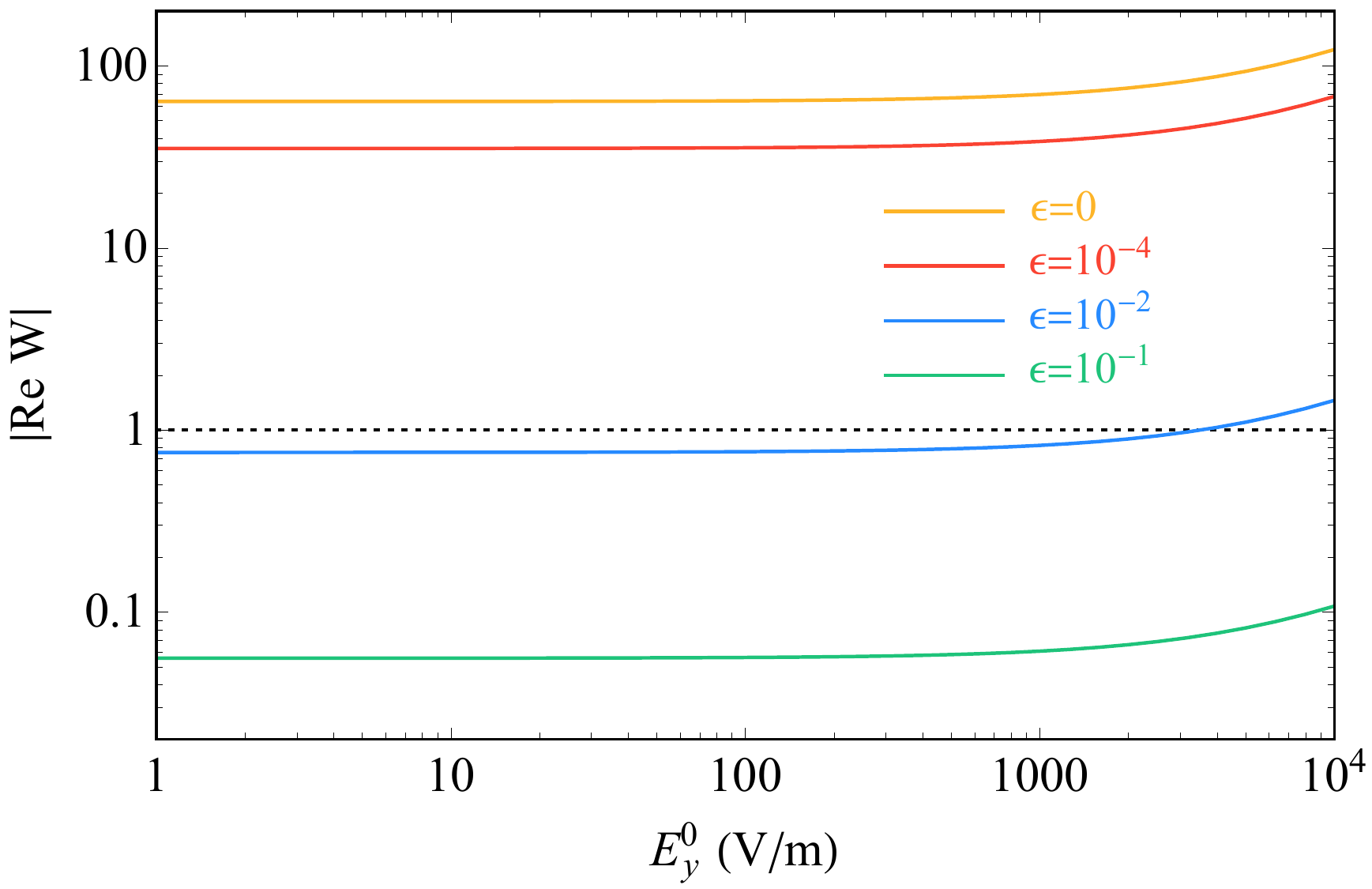}
\end{center}
\vspace{-0.4cm}
\caption{
Left: the weak value $\textrm{Re}W$ in Eq.~\eqref{eq: weakvalue} as a function of $\delta$ with 
$E_{y}^0 = 10$\,V/m fixed.
Right: the weak value $\textrm{Re}W$ as a function of $E_y^0$ with 
$\delta = 0.01 \pi$ fixed.
In both panels, we take 
$T = 10^{-3}$\,sec, 
$E_x^0 = 10^7$\,V/m, 
and  $d_{\rm n}=10^{-26}\,e$\,cm, and vary $\epsilon=0,\,10^{-4},\,10^{-2}$, and $10^{-1}$. 
The black dotted line corresponds to $|\textrm{Re}W|=1$, 
which is the eigenvalue of $\hat \sigma_x$.
  }
\label{fig:lead1}
\end{figure*}

As we will show in the next section,
choosing suitable input parameters
such like $\epsilon,\,\delta,\, T,$ and $E_y^0/E_x^0$,
the weak value $\textrm{Re}W$ 
can be significantly amplified, and it is just the weak value amplification.

Although we analytically obtain
the expectation value of the deviation 
of the neutron position from $x=0$ as $\Delta_x^W$ up to corrections of $\mathcal{O}\left((g_{\mu}\alpha T)^2\right)$, 
 we will also give the full-order result in Appendix~\ref{sec:Full}.  
In the leading-order analysis in Eq.~\eqref{eq:exp},
the EDM signal induced by $\chi$ can be enhanced  by large value of $\alpha T d^2$.
We find, however, 
that 
 the EDM signal is {\it not} amplified by the large value of $\alpha T d^2$ in the full-order analysis,  because an additional damping factor appears 
 as we will discuss
 in the next section.

It is important to distinguish the neutron EDM signal from the background shift.
Since the background effect depends on many parameters and  
is complicated in our setup, 
we evaluate it numerically in the next section.  

Before closing this section, 
let us comment on a special setup in which $\hat{\bf E} =0$ with pre- and post-selections.
In such a case, 
$\Delta_x^W =
   (p_{x0}/m_{\rm n}) T$  is predicted.
Therefore, 
the first term Eq.~\eqref{eq:exp}
can be subtracted by using data of a setup where the external electric field is turned off.

\section{Numerical results}
\label{sec:Result}
%
In this section,
we show numerical results with varying many input parameters.
Here and hereafter, 
we assume the following neutron beam velocity: 
\beq
v_{\rm n} = 10^3\,{\rm m}/ {\rm sec}\,,
\quad
n_{p_{x0}} = n_{p_{y0}}=0\,,
\quad n_{p_{z0}}=1\,,
\eeq
and the neutron beam size
\beq
d = 0.1\,{\rm m}\,,
\eeq
where $v_{\rm n}$ and $d$ are reasonable values 
in the J-PARC neutron beam experiment  \cite{Mishima:2009,Nakajima:2017}, while
ideal values of $n_{p_{x0,y0}}$ are taken.

First, we show the weak value $\textrm{Re}W$ in Eq.~\eqref{eq: weakvalue} in Fig.~\ref{fig:lead1}.
In both panel, 
the black dotted lines represent $\textrm{Re}W =\pm 1$, 
which corresponds to the eigenvalues of $\hat{\sigma}_x$.
Hence, the weak value gives amplification of the signals 
for the regions with $|\textrm{Re}W | > 1$.
In the left panel, we investigate $\delta$ and $\epsilon$ dependence, where
$T = 10^{-3}$\,sec, 
$E_x^0 = 10^7$\,V/m, 
$E_{y}^0 =10\,$V/m,
and  $d_{\rm n}=10^{-26}\,e$\,cm are taken.

Here, we classify the weak measurement in the setup.
Taking $\delta = \pi$ ($\alpha =0$ in Ref.~\cite{Aharonov:1988xu}) 
corresponds to an ordinary indirect measurement, 
where the final beam shift is given by the eigenvalue of the spin operator of the EDM direction $\langle \psi | \hat \sigma_x | \psi\rangle $, in addition to the classical motion
[the first term in Eq.~\eqref{eq:exp}].
In this setup, 
$|\psi\rangle $ is set as $| +_y\rangle$ omitting  the impurity $\epsilon$, so that a zero eigenvalue is obtained as 
$\langle +_y | \hat \sigma_x | +_y\rangle  =0$, which implies that
the beam shift occurs only by  the classical motion.
As you can see in the left panel of Fig.~\ref{fig:lead1}, the weak value $W$ is significantly suppressed around $\delta = \pi$.
Also, one can consider a different setup: $\delta = \pi$ with $|\psi\rangle = | +_x\rangle$.
In this case, the beam shift is given by $\langle +_x | \hat \sigma_x | +_x\rangle  =1$ term in addition to the classical motion. The black dotted line in Fig.~\ref{fig:lead1} shows this latter setup.

It is shown that the weak value amplification, $|{\rm Re}W|>1$, occurs 
when $\epsilon \lesssim 10^{-2}$, 
and the amplification is maximized for small $\delta$ regions. 
As you can see, two orders of magnitude amplification is possible by small $\epsilon$ and $\delta$. 
In the right panel, 
 $E_y^0$ dependence of the weak value is investigated, where 
 $\delta = 0.01 \pi$ is fixed.
It is found that $E_y^0$ dependence is negligible.
Note that we also observed that the weak value is insensitive to $T$, $E_{x}^0$, and $d_{\rm n}$ 
for $|g_{\mu}E_x^0 T|\ll 1$ region [see Eq.~\eqref{eq:gET}].
Above results show that
the weak value amplification can be controlled by only the polarization angle $\delta$ and the selection impurity $\epsilon$ in the pre- and post-selections.

{Here, 
we comment about a weak-measurement approximation, 
which is evaluation up to the first order of $\chi$. 
For $d_{\rm n}=10^{-26}~e~{\rm cm}$, 
the parameter $\chi$ is $1.49\times 10^{-7}$, and the real part of the weak value ${\rm Re}\,W$ is smaller than $10^3$ according to the left panel of Fig.~\ref{fig:lead1}.
Consequently, our evaluation based on the weak-measurement approximation is valid.}

\begin{figure*}[tp]
\begin{center}
\includegraphics[width=8cm]{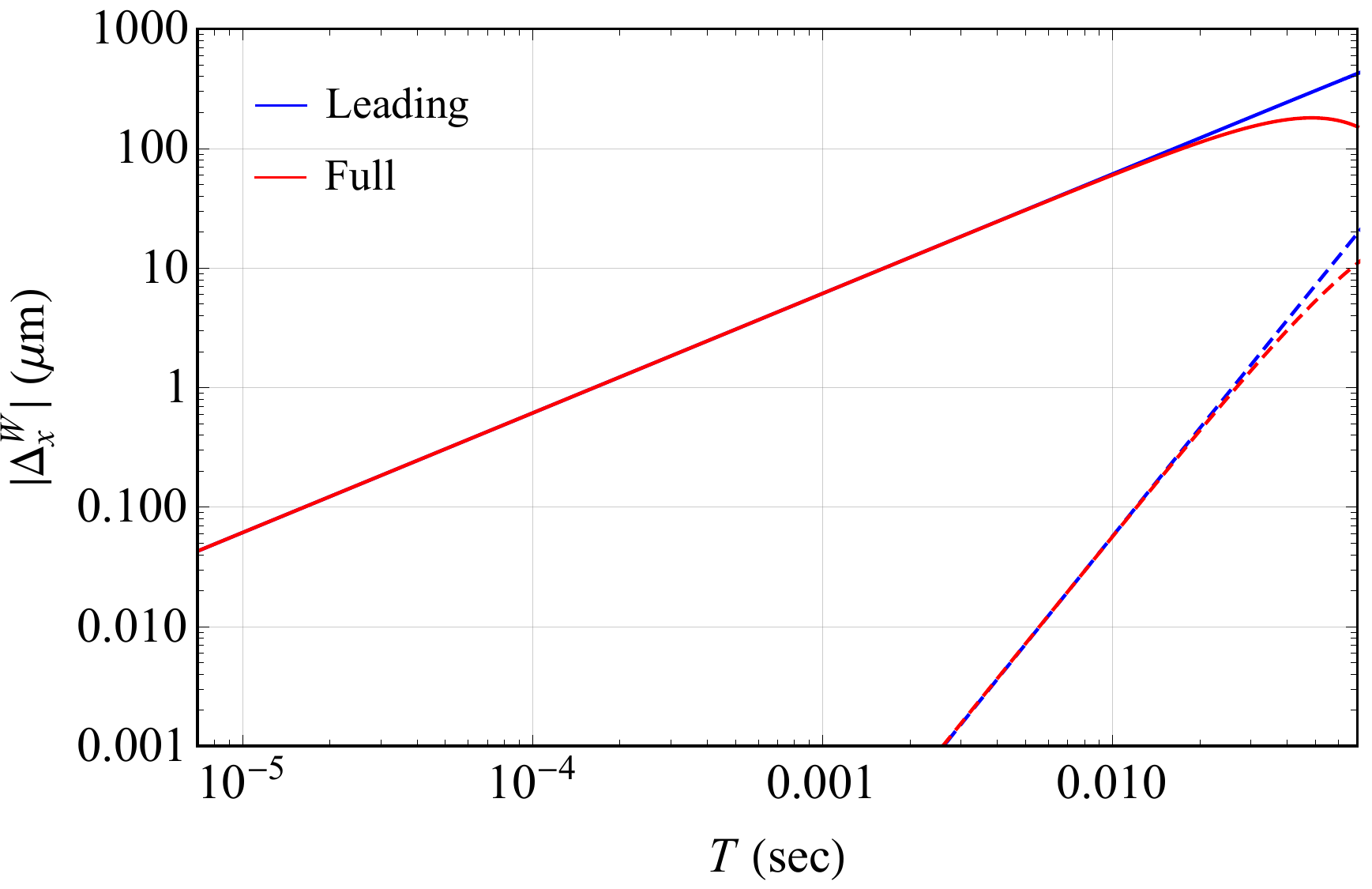}
\quad
\includegraphics[width=8cm]{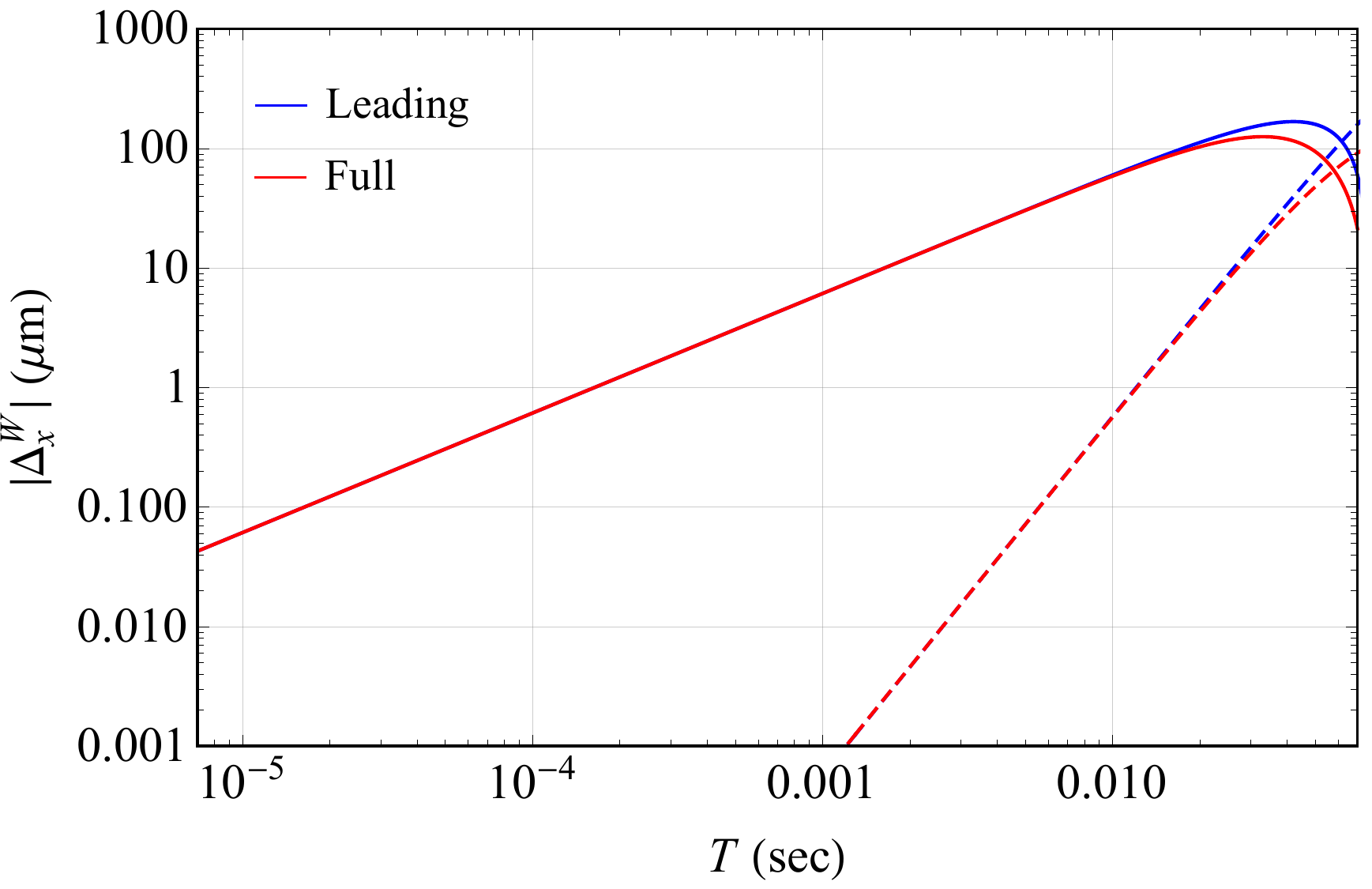}
\end{center}
\vspace{-0.4cm}
\caption{
The expected position shift of the center-of-mass of the neutron bunch  $\Delta_x^W$ is shown as a function of $T$.
The solid lines stand for the EDM signal parts $\Delta_x^W (\textrm{EDM})$ with $d_{\rm n}=10^{-25}\,e$\,cm, while the dashed lines are for
the background effects   $\Delta_x^W (\textrm{BG})$.
The blue and red lines represent the leading-order calculations and the full-order ones. 
In the left and right panels, we take 
$E_x^0 = 10^6$ and $10^7$\,V/m, respectively.  
In both panels,  
$\epsilon=0$,
$\delta= 10^{-3}$,
$E_y^0 =10$\,V/m,
and 
$\alpha = 10^8$\,V/m$^2$ are taken. 
}
\label{fig:ful_lo}
\end{figure*}

Next, we compare the leading-order approximation with respect to $g_{\mu}\alpha T$,  which is shown in the previous section, 
with the full-order analysis.
Since equations for the full-order analysis are lengthy, we put them
on Appendix~\ref{sec:Full}.
From the second term of Eq.~(\ref{eq:exp}),
the expected position shift of the center-of-mass of the neutron bunch  is nearly proportional to $\alpha  T d^2$ 
within the leading-order approximation.
Thus, it is expected that
one can amplify the EDM signal 
by adopting a large value of $\alpha  T d^2$.
This fact is, however, incorrect in the full-order analysis. 
As shown in Appendix~\ref{sec:Full}, 
the full-order results include a damping factor
$\exp\left\{-(g_{\mu}\alpha T)^2 [d^2 +(T/2m_{\rm n})^2/4d^2]/2\right\}$ with respect to $\alpha  T d^2$ 
appeared in Eqs.~(\ref{eq:bil1}) and \eqref{eq: f2}.
Since this damping factor becomes significant for a region of
\beq
g_\mu \alpha T d > 1\,,
\eeq
the EDM signal cannot be amplified  by the large value of $\alpha  T d^2$.
This factor comes from a Gaussian integral,
\beq
\int dx \frac{1}{\sqrt{2 \pi d^2}} e^{i (g_\mu \alpha T) x} e^{- \frac{x^2}{2 d^2}} =
e^{- \frac{1}{2}(g_\mu \alpha T)^2 d^2}\,,
\eeq
and makes the transition probability in Eq.~(\ref{eq:trans}) finite
even when
$\delta\to 0$
and 
the ideal experimental setup limit are taken: 
$E^0_y/E_x^0\to 0$, $n_{p_{x0,y0}}\to 0$,
and $\epsilon\to 0$.

In Fig.~\ref{fig:ful_lo}, 
we show the expected position shift of the center-of-mass of the neutron bunch $\Delta_x^W$  as a function of $T$.
In both panels,
the solid lines stand for the EDM signal parts $\Delta_x^W (\textrm{EDM})$,  which are proportional to $\chi$,   with $d_{\rm n}=10^{-25}\,e$\,cm.
On the other hand, 
the dashed lines are for
the background effects  $\Delta_x^W (\textrm{BG})$, 
which are  $\chi$ independent. 
The blue and red lines correspond to the leading-order calculations and the full-order ones, respectively. 
We take 
$\epsilon=0$,
$\delta= 10^{-3}$,
$E_y^0 =10$\,V/m,
$\alpha = 10^8$\,V/m$^2$, and $E_x^0 = 10^6$ ($10^7$)\,V/m for the left (right) panel.

It is shown that,  
the leading- and the full-order calculations are well consistent with each other in small $T$ regions.
On the other hand, 
$\Delta_x^W$ is significantly suppressed for large $T $ regions.
This figures also show that the background effects in small $T$ region are smaller than the EDM signal contributions by several orders of magnitude,
which has been shown analytically in the previous section. 
In order to suppress
the background effects, 
we adopt $T=0.01\,{\rm sec}$ in following estimations.
We also find that the EDM signal contribution is insensitive to $E_x^0$, while the background effect is sensitive. 
Note that the background effect is also scaled by $E_y^0$ [see Eq.~\eqref{eq:DeltaBG}].

\begin{figure}[t]
\begin{center}
\includegraphics[width=8cm]{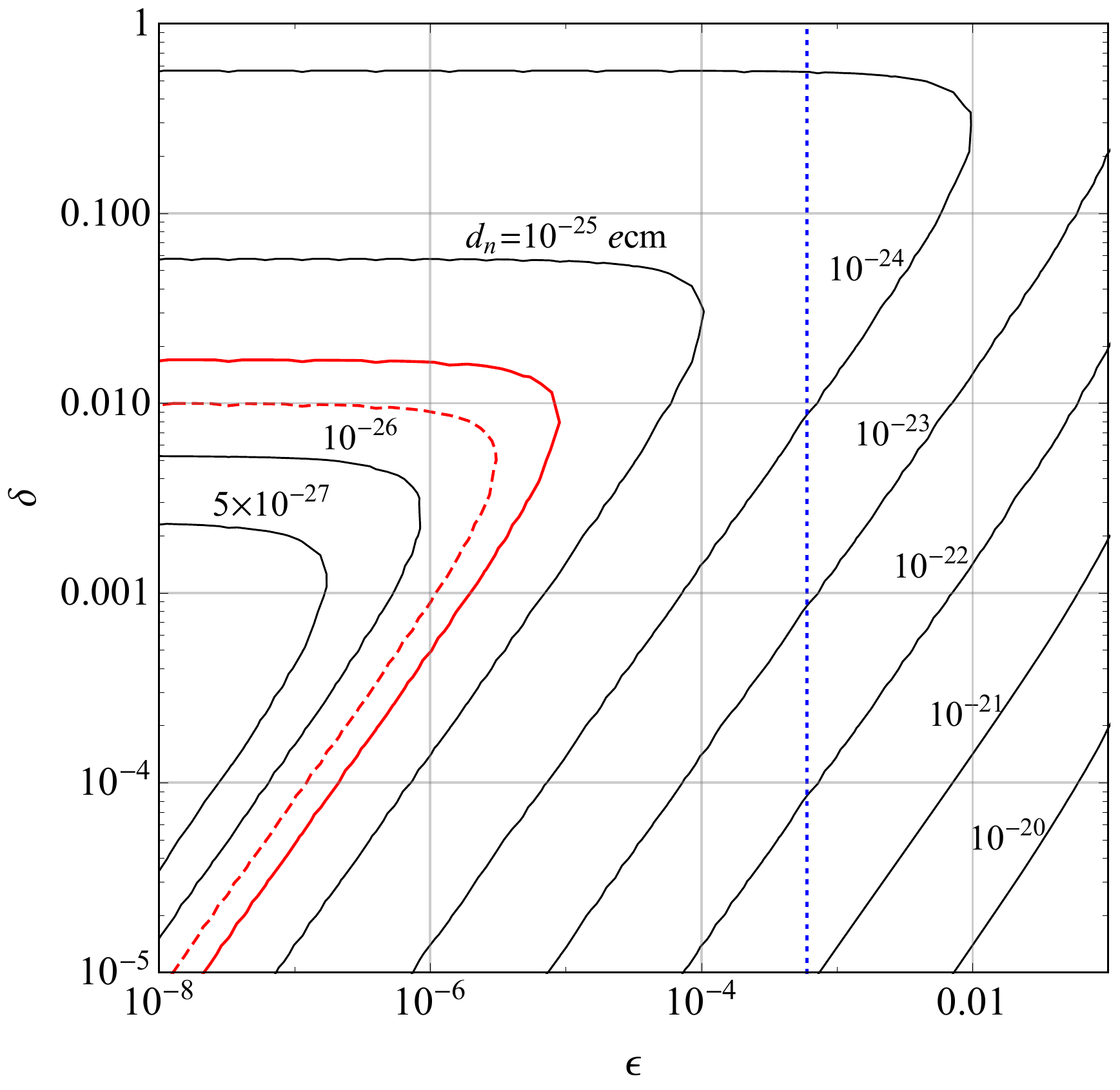}
\end{center}
\vspace{-0.4cm}
\caption{
The potential sensitivity to the neutron EDM as functions of
 the selection impurity $\epsilon$ and the polarization angle $\delta$ 
in the pre- and post-selections.
Here, we take
$T=0.01$\,sec,
$E_x^0 = 10^7$\,V/m,
$E_y^0 = 10$\,V/m, 
and 
$\alpha = 10^8$\,V/m$^2$.
  }
\label{fig:curr1}
\end{figure}

Finally, 
we show a potential sensitivity of the weak measurement that can probe 
the neutron EDM signal.
In this setup, 
the neutron EDM can be probed 
by precise measurement of $\Delta_x^W$.
In current technology, 
it is possible to measure 
the neutron position by several methods 
 with spatial resolutions of
 $100\,{\rm nm}$ \cite{Naganawa:2018xyo},
$1\,\mu {\rm m}$ \cite{Jenke:2013},
$2\,\mu {\rm m}$ \cite{HUSSEY20179}, 
$5\,\mu {\rm m}$ \cite{MATSUBAYASHI2010637,Trtik_2016},
$22\,\mu {\rm m}$,  \cite{Shishido2018}, and $50\,\mu {\rm m}$
\cite{Kuroda:1989pq}.
These spatial resolutions determine
the potential sensitivity to the neutron EDM signal.
The detector schemes of Refs.~\cite{MATSUBAYASHI2010637,Trtik_2016,Shishido2018,Kuroda:1989pq} are examined for the neutron beam, 
and a thermal neutron is examined in Ref.~\cite{HUSSEY20179}.
On the other hand, the ones of Refs.~\cite{Jenke:2013,Naganawa:2018xyo} are examined for the UCN, but they can also be  utilized for cold neutron (neutron beam) with small detection efficiency \cite{mishima}.
Recently, the detector scheme of Ref.~\cite{Naganawa:2018xyo} has been improved~\cite{mishima}, and the detection efficiency becomes $\mathcal{O}(1\%)$ with spatial resolution of $1\text{--}2\,\mu {\rm m}$ for the cold neutron.
Although the emitted neutron beam size is significantly larger than the spatial resolution of the detector, 
it would not raise a matter. 
Rather, the spatial resolution should be compared with the statistical uncertainty of the beam size, which will be discussed in the end of this section.

By requiring a condition $|\Delta_x^W|> 1\,\mu {\rm m}$ as a reference value, 
we show the sensitivity to the neutron EDM  in  Fig.~\ref{fig:curr1},
where $\Delta_x^W $ includes both the EDM signal and the background shift, and the full-order formalism is used.
Here, notice that 
the information necessary for this analysis 
is only a time-averaged shift of the center-of-mass of the neutrons passing the post-selection.
The sensitivity is shown as a contour on the $\epsilon$--$\delta$ plane, here we take
$T=0.01$\,sec,
$E_x^0 = 10^7$\,V/m,
$E_y^0 = 10$\,V/m, 
and 
$\alpha = 10^8$\,V/m$^2$.
Based on parameters in Refs.~\cite{PhysRevC:88:045502,Dress:1976bq} and discussions with an experimentalist at the J-PARC~\cite{mishima}, 
 these parameters are chosen.
 In the Fig.~\ref{fig:curr1}, the red (dashed) line corresponds to
the current (improved) neutron EDM bound in Eq.~\eqref{eq:currentbound} [Eq.~\eqref{eq:currentbound_improved}], 
and the larger $d_{\rm n}$ region is excluded.
Moreover, we find the background effect is negligible on this plane. 

We show that the impurity effect changes the sensitivity drastically, and 
find that the neutron EDM signal can be probed for a very small impurity region, $\epsilon < 10^{-5}$.
It is two orders of magnitude smaller than the current technology, \eg,  
$\epsilon = (6 \pm 1_{\rm stat.} \pm 3_{\rm sys.})\times 10^{-4}  $ \cite{Yoshioka:2011}, which is shown as the vertical blue dotted line in  Fig.~\ref{fig:curr1}:
the setup with $\epsilon = 6 \times 10^{-4}$ and $\delta= 0.1$ $= 5.7^{\circ}$  could probe a region $d_{\rm n} > 3 \times 10^{-25}~e\,{\rm cm}$.

We comment on contributions from nonzero $n_{p_{x0}}$ and $n_{p_{y0}}$ values.
We find that if $n_{p_{x0,y0}}$ are smaller than $10^{-5}$, these effects do not appear in above numerical evaluations.
We also find that the effect of $n_{p_{y0}}$ is more significant than  $n_{p_{x0}}$: $\mathcal{O}(1)$ contributions are produced for $n_{p_{y0}} \sim 10^{-4}$ region.

We also comment on a statistical condition for measuring non-zero $\Delta_{x}^W$,
where we compare the spatial resolution of the detector with the statistical uncertainty of the neutron beam.
In such an experiment, 
one has to reject a null hypothesis 
of the neutron beam following the Gaussian distribution with the average position $0$ and the variance $d^2$. 
If one measures the time-averaged shift of the center-of-mass of the neutrons 
by $\mathcal{N}$ neutrons,
the statistical condition for measuring non-zero $\Delta_{x}^W$
at $n\sigma$ level is expressed as
$ \Delta_{x}^W > n\cdot d/\sqrt{\mathcal{N}}$. 
If one considers a case that a resolution of $1\,\mu$m for $\Delta_{x}^W$ and $d= 0.1\,{\rm m}$, the condition is $\mathcal{N }>  10^{10} n^2$.
In this setup, the number of neutrons $\mathcal{N}$ is
\beq
\mathcal{N} \simeq \mathcal{N}_0 \cdot 2 \epsilon \cdot \varepsilon_{\rm eff}\,,
\eeq
where $\mathcal{N}_0$ represents the number of emitted neutrons, $2 \epsilon$ is the neutron passing probability at the post-selection in Eq.~\eqref{eq:passing}, and $\varepsilon_{\rm eff}$ is the detection efficiency at the detector.
Then, the statistical condition is 
\beq
\mathcal{N}_0 > \frac{10^{10} n^2}{2 \epsilon \varepsilon_{\rm eff}}\,.
\eeq
Since the $\mathcal{O}(10^9)$  neutrons can be generated per second in the experiment~\cite{Nakajima:2017},
a required time is $t > 5 n^2 /(\epsilon \varepsilon_{\rm eff})$\,sec.
Even if the detection efficiency for the neutron beam is $\mathcal{O}(1\%)$ \cite{Jenke:2013,mishima}, 
the statistical condition for $n=2$ is satisfied by $\mathcal{O}(2000/\epsilon)$ seconds neutron beam.

\section{Conclusions}
\label{sec:conclusion}
%

In this paper,
we proposed a novel approach in a search for the neutron EDM 
by applying the weak measurement, which is independent from the Ramsey method.
Although the relativistic ${\bf E} \times {\bf v}$ effect
provides
a severe systematic uncertainty 
in the neutron EDM experiment in which the neutron beam is used,
we find such a contribution
is numerically irrelevant in the weak measurement.
This is quite unexpected result, and 
we believe that this fact would provide  a new virtue of the weak measurement.

To investigate a potential sensitivity to the neutron EDM search, 
we included the effect from the selection impurities in the pre- and post-selections.
Our study showed that the size of the impurity crucially determines the sensitivity.

We found that
the weak measurement can reach up to $d_{\rm n} > 3 \times 10^{-25}~e\,{\rm cm}$ within the current technology. 
This is one order of magnitude less sensitive that the current neutron EDM bound,
where the UCNs based on the Ramsey method are used.

In addition, our approach 
could provide a new possibility 
to measure the weak value of the neutron spin polarization
from the background effect.
This fact makes our study fascinating
in the point of view of the quantum mechanics.

{
The detailed study  about the Fisher information
based on Ref.~\cite{Harris_2017} would be one of future directions of this study, 
 where one can explore 
 whether the weak-value amplification in the neutron EDM measurement outperforms the conventional Ramsey method one. 
Also, the perturbative effects from the Gaussian beam profile~\cite{Clark} could give additional systematic error in the weak measurement~\cite{Turek_2015}.
}

Although the small impurity, $\epsilon < 10^{-5}$, 
for probing the neutron EDM is difficult at the present time,
we hope several improvements on the experimental technology, \eg, the sensitivity can be amplified by $\alpha$ and the resolution of $\Delta_x^W$ measurement,
and anticipate that this kind of experiment will be performed in future.


\section*{Acknowledgments}
We are grateful to
Marvin Gerlach
for collaboration in the early stage of this project. 
We would like to thank 
Joseph Avron, Yuji Hasegawa, Masataka Iinuma, Oded Kenneth, and  Izumi Tsutsui 
for worthwhile discussions of the weak measurements. 
We also thank Kenji Mishima 
for helpful discussions about current neutron beam experiments.
We greatly appreciate many valuable conversations with our colleagues, 
Gauthier Durieux,
Masahiro Ibe, 
Go Mishima, 
Yael Shadmi,
Yotam Soreq,
and Yaniv Weiss. 
D.U.\ would like to thank all the members of Institute for Theoretical Particle Physics (TTP) at Karlsruhe Institute of Technology, especially
Matthias Steinhauser,
for kind hospitality during the stay.
D.U.\ also
appreciates being invited to conferences of the weak value and weak measurement at KEK.
The work of T.K.\ is supported by 
the Israel Science Foundation (Grant No.~751/19) and by the Japan Society for the Promotion of Science (JSPS) KAKENHI Grant Number 19K14706.

\onecolumngrid
\appendix 

\section{External electric field with gradient}
\label{App:A}
%
In this appendix,
we consider a general setup 
of the external electric field with spatial gradients.
Let us consider an external electric field with gradients 
$\partial \hat{\bf E}/{\partial x}=(\alpha_x,\,\alpha_y,\,0)$: 
\begin{align}
    \hat{\bf E}(x)=\left(E_x^0 +\alpha_x x,\,E_y^0 +\alpha_y x,\,0 \right)\,,
\end{align}
where $E_{x,y}^0$ and $\alpha_{x,y}$ are constants.
Note that the setup is insensitive to $E_z$ component and its relativistic effect. 

When one considers a rotation of the coordinate
that all spatial dependence go to single electric filed component, 
the following $\hat{\bf E}$ with a coordinate ${\bf x}'$ are obtained: 
\begin{align}
    \hat{\bf E}'(x',y')^{T} &=R \hat{\bf E}^{T}
    =\frac{1}{\sqrt{\alpha_x^2 +\alpha_y^2}}
    \begin{pmatrix}
    \alpha_x E_x^0 +\alpha_y E_y^0+(\alpha_x^2 +\alpha_y^2)x \\
    -\alpha_y E_x^0 +\alpha_x E_y^0 \\
    0
    \end{pmatrix}\\
    &= \begin{pmatrix}
    \frac{1}{\sqrt{\alpha_x^2+\alpha_y^2}}(\alpha_x E_x^0 +\alpha_y E_y^0)+\alpha_x x' -\alpha_y y' \\
    \frac{1}{\sqrt{\alpha_x^2+\alpha_y^2}}(-\alpha_y E_x^0 +\alpha_x E_y^0) \\
    0
    \end{pmatrix}\,,
\end{align}
and
\begin{align}
    {\bf x}'\equiv \begin{pmatrix}
       x'\\
        y' \\
        z'
    \end{pmatrix}=  R {\bf x} =\frac{1}{\sqrt{\alpha_x^2 +\alpha_y^2}}
    \begin{pmatrix}
        \alpha_x x +\alpha_y y\\
        -\alpha_y x +\alpha_x y \\
        z
    \end{pmatrix}\,,
\end{align}
where 
the rotation matrix $R$ is
\begin{align}
    R=
    \begin{pmatrix}
    \cos \theta & \sin\theta & 0\\
    -\sin \theta & \cos \theta & 0\\
    0 & 0 &1
    \end{pmatrix}\,,
    ~~{\rm with}~~\cos\theta=\frac{\alpha_x}{\sqrt{\alpha_x^2 +\alpha_y^2}}\,.
\end{align}
The spatial dependence of $y'$ in $\hat{\bf E}'(x',y')$ 
would provide us 
with a spatial displacement of the neutron  along $y'$ axis.
In order to obtain the experimental setup in Eq.~\eqref{eq:Eapp},
$\alpha_y\ll \alpha_x$ are thus required.
%

\section{Full-order calculation for \texorpdfstring{\mbox{\boldmath$\Delta_x^W$}}{DeltaxW}}
\label{sec:Full}
%
In this appendix,
we give building blocks of the full-order calculation for $\Delta_x^W$ with respect to $g_{\mu}\alpha T$.
The expected position shift of the center-of-mass of the neutron bunch is defined as [see Eqs.~\eqref{eq:trans}, \eqref{eq:trans2}, and \eqref{eq:exp} for the leading-order analysis]:
\beq
\Delta_x^W =  \frac{{\rm Tr}\left[\left(\hat{x} + \frac{T}{2 m_{\rm n}}\hat{p}_x  \right)\hat{\rho}^{\rm total}_{\rm fin} \right]}
{ {\rm Tr}\left[\hat{\rho}^{\rm total}_{\rm fin}\right]}\,,
\eeq
with
\begin{align}
    {\rm Tr}\left[\hat{\rho}^{\rm total}_{\rm fin}\right]&=  {\langle G_{p_{x0}}|{\rm Tr}_S \left[\hat{\rho}^S_{\rm fin} e^{-iT \hat{H}}\hat{\rho}^S_{\rm ini}e^{i T\hat{H}} \right]|G_{p_{x0}} \rangle}\,,\label{eq: ful1}
    \\
    {\rm Tr}\left[\left(\hat{x} + \frac{T}{2 m_{\rm n}}\hat{p}_x  \right)\hat{\rho}^{\rm total}_{\rm fin} \right]&= {\langle G_{p_{x0}}|\left(\hat{x}+\frac{T}{2 m_{\rm n}}\hat{p}_x  \right){\rm Tr}_S \left[\hat{\rho}^S_{\rm fin} e^{-iT \hat{H}}\hat{\rho}^S_{\rm ini}e^{i T\hat{H}} \right] |G_{p_{x0}}\rangle}\,.\label{eq: ful2}
\end{align}
In the full-order analysis, we discard only $\mathcal{O}(\chi^2)$ contributions. Using Eqs.~\eqref{eq:H0Hchi}--\eqref{eq:Hchi1} for the time evolution operator,
${\rm Tr}_S [\hat{\rho}^S_{\rm fin} e^{-iT \hat{H}}\hat{\rho}^S_{\rm ini}e^{i T\hat{H}} ]$ is expanded as
\begin{align}
    {\rm Tr}_S\left[\hat{\rho}^{S}_{\rm fin}e^{-iT \hat{H}}\hat{\rho}^S_{\rm ini}e^{iT \hat{H}} \right]=&
    e^{- i T \frac{\hat{p}^2_x}{2 m_{\rm n}}} \biggl\{  
 -\frac{1}{2 \hat{E}_x^2} \bigg[\hat{E}_x^2 (-1 + (1 - 2 \epsilon)^2 \cos(\delta)) \cos^2(
     \hat{E}_x g_{\mu} T) - (E^0_y n_{p_{x0}} - \hat{E}_x n_{p_{y0}})^2 \sin^2(
     \hat{E}_x g T)\notag
     \\
    & - (\hat{E}_x^2 + (E^0_y)^2) n_{p_{z0}}^2 \sin^2(
     \hat{E}_x g_{\mu} T) - (1 - 2 \epsilon)^2 (E^0_y n_{p_{x0}} - \hat{E}_x n_{ p_{y0}})^2 \cos(\delta) \sin^2(
     \hat{E}_x g_{\mu} T)\notag
     \\
     & - (1 - 2 \epsilon)^2 (-\hat{E}_x + E^0_y) (\hat{E}_x + E^0_y) n_{p_{z0}}^2 \cos(\delta) \sin^2(
     \hat{E}_x g_{\mu} T)\notag
\\     
     & -
   2 (1 - 2 \epsilon)^2 \hat{E}_x (E^0_y n_{p_{x0}} - \hat{E}_x n_{ p_{y0}}) n_{p_{z0}} \sin(\delta) \sin^2(
     \hat{E}_x g_{\mu} T) + (1 - 2 \epsilon)^2 \hat{E}_x E^0_y n_{p_{z0}} \sin(\delta) \sin(2 \hat{E}_x g_{\mu} T)\bigg]\notag
     \\
  & -\chi\frac{1}{2 \hat{E}_x^2}
  \bigg[-2 \hat{E}_x E^0_y n_{p_{z0}} \sin^2(\hat{E}_x g_{\mu} T) - 
    2 (1 - 2 \epsilon)^2 \hat{E}_x E^0_y n_{p_{z0}} \cos(\delta) \sin^2(
      \hat{E}_x g_{\mu} T)\notag
      \\
      & + (1 - 2 \epsilon)^2 \hat{E}_x^2 \sin(\delta) \sin(2 \hat{E}_x g_{\mu} T)\bigg]  \biggr\} e^{+ i T \frac{\hat{p}^2_x}{2 m_{\rm n}}} 
    +\mathcal{O}(\chi^2)\,,
    \label{eq:fulspin}
\end{align}
where $\hat{E}_x$ is defined as $\hat{E}_x = E_x^0 + \alpha \left(\hat{x} +T \hat{p}_x/2m_{\rm n} \right)$.
 To calculate Eqs.~(\ref{eq: ful1}) and (\ref{eq: ful2}), the following building blocks are needed:
  \begin{align}
   &  {\langle G_{p_{x0}}|e^{-i T\frac{\hat{p}_x^2}{2 m_{\rm n}}}\cos^2\left( \hat{E}_x g_{\mu}T\right) e^{+i T\frac{\hat{p}_x^2}{2 m_{\rm n}}} |G_{p_{x0}}\rangle}=\frac{1}{4}\left[f_1(2 g_{\mu})+f_1(-2 g_{\mu})+2 \right]\,,
   \\
   &  {\langle G_{p_{x0}}|e^{-i T\frac{\hat{p}_x^2}{2 m_{\rm n}}}\sin^2\left( \hat{E}_x g_{\mu}T\right) e^{+i T\frac{\hat{p}_x^2}{2 m_{\rm n}}} |G_{p_{x0}}\rangle}=-\frac{1}{4}\left[f_1(2 g_{\mu}) +f_1 (-2 g_{\mu})-2 \right]\,,
   \\
   & {\langle G_{p_{x0}}|e^{-i T\frac{\hat{p}_x^2}{2 m_{\rm n}}}\frac{1}{\hat{E}_x^2}\sin^2 \left(\hat{E}_x g_{\mu}T\right) e^{+i T\frac{\hat{p}_x^2}{2 m_{\rm n}}} |G_{p_{x0}}\rangle}=T^2 \int_0^{g_{\mu}} dg_1 \int_0^{g_1}dg_2 \left[f_1(2 g_2)+f_1 (-2 g_2)  \right]\,,
   \\
   & {\langle G_{p_{x0}}|e^{-i T\frac{\hat{p}_x^2}{2 m_{\rm n}}}\frac{1}{\hat{E}_x}\sin^2\left( \hat{E}_x g_{\mu}T\right) e^{+i T\frac{\hat{p}_x^2}{2 m_{\rm n}}} |G_{p_{x0}}\rangle}=\frac{T}{2i}\int_0^{g_{\mu}} dg_1 \left[f_1(2 g_1)-f_1(-2 g_1) \right]\,,
   \\
   & {\langle G_{p_{x0}}|e^{-i T\frac{\hat{p}_x^2}{2 m_{\rm n}}}\sin \left(2\hat{E}_x g_{\mu}T\right) e^{+i T\frac{\hat{p}_x^2}{2 m_{\rm n}}} |G_{p_{x0}}\rangle}=\frac{1}{2i}\left[f_1(2 g_{\mu}) -f_1(-2 g_{\mu}) \right]\,,
   \\
   & {\langle G_{p_{x0}}|e^{-i T\frac{\hat{p}_x^2}{2 m_{\rm n}}}\frac{1}{\hat{E}_x}\sin \left(2\hat{E}_x g_{\mu}T\right) e^{+i T\frac{\hat{p}_x^2}{2 m_{\rm n}}} |G_{p_{x0}}\rangle}= T \int_0^{g_{\mu}} dg_1 \left[f_1(2 g_1) +f_1(-2g_1) \right]\,,
   \\
   & {\langle G_{p_{x0}}|\left(\hat{x} +\frac{T}{2 m_{\rm n}}\hat{p}_x\right) e^{-i T\frac{\hat{p}_x^2}{2 m_{\rm n}}}\cos^2 \left(\hat{E}_x g_{\mu}T\right) e^{+i T\frac{\hat{p}_x^2}{2 m_{\rm n}}} |G_{p_{x0}}\rangle}=\frac{1}{4} \left[f_2(2 g_{\mu}) + f_2 (-2 g_{\mu}) +\frac{T}{2 m_{\rm n}}p_{x0} \right]\,,
   \\
  & {\langle G_{p_{x0}}|\left(\hat{x} +\frac{T}{2 m_{\rm n}}\hat{p}_x\right) e^{-i T\frac{\hat{p}_x^2}{2 m_{\rm n}}}\sin^2 \left(\hat{E}_x g_{\mu}T\right) e^{+i T\frac{\hat{p}_x^2}{2 m_{\rm n}}} |G_{p_{x0}}\rangle}=-\frac{1}{4} \left[f_2 (2 g_{\mu})+f_2 (-2 g_{\mu})-\frac{T}{2 m_{\rm n}}p_{x0}\right]\,,
  \\
& {\langle G_{p_{x0}}|\left(\hat{x} +\frac{T}{2 m_{\rm n}}\hat{p}_x\right) e^{-i T\frac{\hat{p}_x^2}{2 m_{\rm n}}}\frac{1}{\hat{E}_x^2}\sin^2\left( \hat{E}_x g_{\mu}T\right) e^{+i T\frac{\hat{p}_x^2}{2 m_{\rm n}}} |G_{p_{x0}}\rangle}=T^2 \int_0^{g_{\mu}}d g_1 \int_0^{g_1} dg_2 \left[f_2 (2 g_2) +f_2 (-2 g_2) \right]\,,
\\
& {\langle G_{p_{x0}}|\left(\hat{x} +\frac{T}{2 m_{\rm n}}\hat{p}_x\right) e^{-i T\frac{\hat{p}_x^2}{2 m_{\rm n}}}\frac{1}{\hat{E}_x}\sin^2\left( \hat{E}_x g_{\mu}T\right) e^{+i T\frac{\hat{p}_x^2}{2 m_{\rm n}}} |G_{p_{x0}}\rangle}=\frac{T}{2 i}\int_0^{g_{\mu}}dg_1 \left[f_2 (2 g_1)-f_2 (-2 g_1) \right]\,,
\\
& {\langle G_{p_{x0}}|\left(\hat{x} +\frac{T}{2 m_{\rm n}}\hat{p}_x\right) e^{-i T\frac{\hat{p}_x^2}{2 m_{\rm n}}}\sin \left(2\hat{E}_x g_{\mu}T\right) e^{+i T\frac{\hat{p}_x^2}{2 m_{\rm n}}} |G_{p_{x0}}\rangle}=\frac{1}{2 i}\left[f_2 (2 g_{\mu})-f_2 (-2 g_{\mu}) \right]\,,
\\
& {\langle G_{p_{x0}}|\left(\hat{x} +\frac{T}{2 m_{\rm n}}\hat{p}_x\right) e^{-i T\frac{\hat{p}_x^2}{2 m_{\rm n}}}\frac{1}{\hat{E}_x}\sin \left(2\hat{E}_x g_{\mu}T\right) e^{+i T\frac{\hat{p}_x^2}{2 m_{\rm n}}} |G_{p_{x0}}\rangle}=T \int_0^{g_{\mu}}dg_1 \left[f_2(2 g1) +f_2 (-2 g_1) \right]\,,
\label{eq:bile}
\end{align}
with
\begin{align}
   & f_1(g_{\mu})\equiv  {\langle G_{p_{x0}}|e^{-i T\frac{\hat{p}_x^2}{2 m_{\rm n}}}e^{i \hat{E}_x g_{\mu}T}e^{+i T\frac{\hat{p}_x^2}{2 m_{\rm n}}} |G_{p_{x0}}\rangle} 
   \notag
   \\
   &\quad \quad ~~ ={e^{i\left(g_{\mu}E_x^0 T -  g_{\mu}\alpha T \frac{T}{2 m_{\rm n}}p_{x0}\right)} e^{-\frac{(g_{\mu}\alpha T)^2}{2}\left[d^2 +\frac{1}{4d^2}\left(\frac{T}{2 m_{\rm n}}\right)^2 \right]}}\,,\label{eq:bil1}
   \\
   & f_2 (g_{\mu})\equiv  {\langle G_{p_{x0}}|e^{-i T\frac{\hat{p}_x^2}{2 m_{\rm n}}}\left(\hat{x}+\frac{T}{2 m_{\rm n}}\hat{p}_x\right) e^{i \hat{E}_x g_{\mu}T}e^{+i T\frac{\hat{p}_x^2}{2 m_{\rm n}}} |G_{p_{x0}}\rangle}
   \notag 
   \\
 &\quad \quad ~~  ={\left\{-\frac{T}{2 m_{\rm n}}(p_{x0}+g_{\mu}\alpha T)+i g_{\mu}\alpha T  \left[d^2 +\frac{1}{4d^2}\left(\frac{T}{2 m_{\rm n}}\right)^2  \right]\right\} e^{i\left(g_{\mu}E_x^0 T -g_{\mu}\alpha T \frac{T}{2 m_{\rm n}}p_{x0}\right)} e^{-\frac{(g_{\mu}\alpha T)^2}{2}\left[d^2 +\frac{1}{4d^2}\left(\frac{T}{2 m_{\rm n}}\right)^2 \right]}}
   \,.\label{eq: f2}
  \end{align}
Combining Eqs.~(\ref{eq: ful1})--(\ref{eq: f2}),
one can numerically calculate the expected position shift of the center-of-mass of the neutron bunch at the full order.
See \eg, Refs.~\cite{Lee_2014,kawanaueda} for detailed discussions of the damping factors in Eqs.~(\ref{eq:bil1}) and (\ref{eq: f2}).     

\twocolumngrid

\bibliography{ref}

\end{document}